\documentclass[onecolumn, 10pt]{article}

\usepackage[utf8]{inputenc}
\usepackage[T1]{fontenc}
\usepackage{mathptmx}
\usepackage{bm}
\usepackage{xcolor}
\usepackage{amsmath}
\usepackage{graphicx}
\usepackage{authblk}
\usepackage[a4paper, total={6in, 9in}]{geometry}

\def\Tr#1{{\ensuremath{\text{Tr}{(#1)}}}}

\def\tensor#1{{\ensuremath{{\bm{#1}}}}}
\def\half{{\textstyle \frac{1}{2}}}

\usepackage{comment}
\def\bH {\ensuremath{\bar{H} }}
\def\bD {\ensuremath{\bar{D} }}
\def\bk {\ensuremath{\bar{\kappa} }}

\begin{document}

\title{Curvature-driven instabilities in thin active shells}

\author{
Andrea Giudici$^{1}$, John S. Biggins$^{1}$}

\affil{$^{1}$ Department of Engineering, University of Cambridge, Trumpington St., Cambridge CB21PZ, U.K.}

\maketitle

\begin{abstract}
Spontaneous material shape changes, such as swelling, growth or thermal expansion, can be used to trigger  dramatic elastic instabilities in thin shells. These instabilities originate in geometric incompatibility between the preferred extrinsic and intrinsic curvature of the shell, which may be modified by active deformations through the thickness and in plane respectively. Here, we solve the simplest possible model of such instabilities, which assumes the shells are shallow, thin enough to bend but not stretch, and subject to homogeneous preferred curvatures. We consider separately the cases of zero, positive and negative Gaussian curvature. We identify two types of super-critical \textit{symmetry breaking} instability, in which the shell's principal curvature spontaneously breaks discrete up-down symmetry and continuous planar isotropy respectively. These are then augmented by  \textit{inversion} instabilities, in which the shell jumps sub-critically between up/down broken symmetry states, and  \textit{rotation} instabilities, in which the curvatures rotate by 90 degrees between states of broken isotropy without release of energy. Each instability has a thickness independent threshold value for the preferred extrinsic curvature proportional to the square-root of Gauss curvature. Finally, we show that the threshold for the isotropy-breaking instability is the same for deep spherical caps, in good agreement with recently published data. 
\end{abstract}

\section{Introduction}
Elastic structures under load are often susceptible to dramatic instabilities such as the buckling of a column under compression\cite{Euleroo,yoo2011stability}, or the collapse of a vessel under pressure \cite{zoleypressure,Hutchinson2016}. Traditionally, engineers have studied instabilities to avoid them as they lead to softening, fracture and failure. However, recently, the study of instabilities has been revitalised by the growing interest in soft materials such as gels, elastomers and biological tissues. Soft materials can undergo large strains without failing, meaning they survive instabilities and are thus capable of dramatic shape changes \cite{kochmann2017exploiting}.
Furthermore, many  soft materials can undergo active strain deformations, including  swelling gels \cite{hirokawa1984volume,klein2007shaping,kim2012designing,na2016grayscale, gladman2016biomimetic}, growth in biological tissues \cite{thompson1942growth, savin2011growth, shyer2013villification, goriely2017mathematics}, uniaxial contraction in liquid crystal elastomerss \cite{de2012engineering, ware2015voxelated, aharoni2018universal, barnes2019direct} and biological muscles, and inflation in barromorphs \cite{siefert2019bio, warner2020inflationary} and other pneumatic systems\cite{mosadegh2014pneumatic}. When these active strain are homogeneous across the material, one obtains simple shape changes. However, when they are inhomogeneous, they are also often geometrically incompatible, meaning that there is no corresponding displacement field that allows the system to fully relax, leading to internal stresses even in the energy minimising state.

Such incompatible spontaneous deformations can trigger spontaneous elastic instabilities, in which the system suddenly adopts a new and more complex shape. For example, a growing layer on a soft substrate can wrinkle due to the inter-facial mismatch between the length of the layer and the substrate it is attached to\cite{bowden1998spontaneous,jiang2007finite,sultan2008buckling,huang2005nonlinear,audoly2007buckling}. Biology uses such growth-induced wrinkling to form brain folds \cite{brain,brainpnas}, villi \cite{villi}, and gut loops \cite{loop}. 
In thin shell-like structures, incompatible active deformations generate instabilities such as the snap of a Venus fly-trap \cite{forterre2005venus}, the folding of a pollen grain \cite{Katifori2010,Couturier2013,Bozic2020} and the bistability of a cyclists snap-band \cite{seffen1999deployment}. These different shell instabilities have been addressed individually, but a clear and intuitive overview of their behaviour is still lacking. Here, we make a start by presenting a complete and simple treatment of incompatibility driven instabilities and states in thin and shallow shells, highlighting the different behaviours of positive, zero and negative Gauss curvature geometries. 

\begin{figure}[t]
    \centering
 \includegraphics[width = \textwidth]{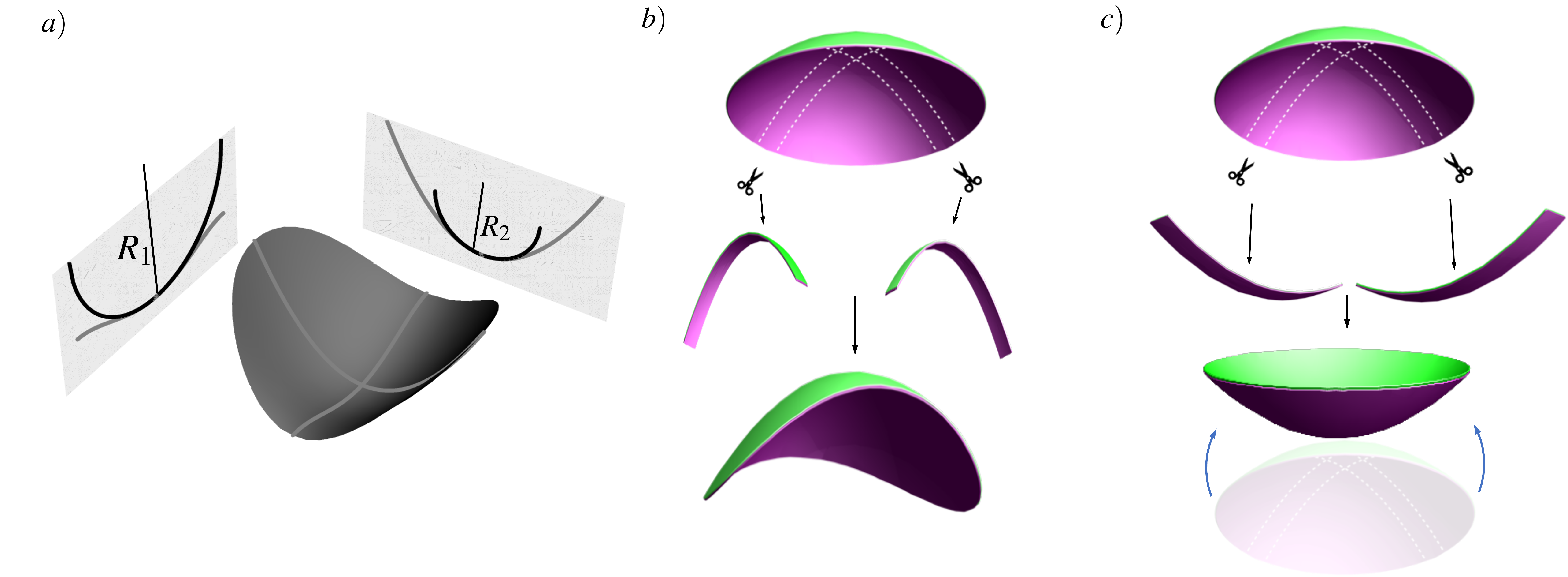}
    \caption{a) Principal curvatures of radius $R_1$ and $R_2$ at a point on a shell. b) A spherical cap subject to swelling of the top (green) layer. Cut-out strips want to curve more than what allowed by the geometry of the spherical cap, leading to the loss of rotational symmetry and folding of the shell. c) A spherical cap subject to swelling of the bottom (purple) layer. Cut-out strips want to curve in the opposite direction as opposed to what allowed by the geometry of the spherical cap. Excessive opposite bend can lead to snap through inversion of the system (bottom image).}
    \label{fig:intro}
\end{figure}

Due to their slenderness, thin shells can be represented by a 2D surface. Any surface in 3D euclidean space can be described by its intrinsic and extrinsic properties. The intrinsic geometry is characterised by distances between points on the surface, captured by the metric tensor $\tensor{a}$.
The extrinsic geometry is locally described by the two principal curvatures $\kappa_1$ and $\kappa_2$, identified by inscribing the largest and smallest possible circles tangent to the surface, as shown in Fig.\ \ref{fig:intro}a). Mathematically, aligning the coordinate system with the curvature directions, means the curvature is described by the tensor $\tensor{\kappa}=\text{diag}(\kappa_1,\kappa_2)$.
However, crucially, $\tensor{a}$ and $\tensor{\kappa}$ cannot be chosen independently --- in any real surface, they must satisfy geometric compatibility conditions. Most famously, they must satisfy Gauss's theorema egregium,stating that the Gauss curvature $K=\det{
\tensor{\kappa}}=\kappa_1 \kappa_2$ is an intrinsic property of the surface and can be computed directly from the metric \cite{kuhnel2015differential}. 

In physical shells, active shape changes can separately modify the locally preferred metric $\bar{\tensor{a}}$ by changing in-plane distances \cite{klein2007shaping,na2016grayscale, gladman2016biomimetic,de2012engineering, ware2015voxelated, aharoni2018universal, barnes2019direct,siefert2019bio,van2017growth,warner2020topographic}, and the locally preferred curvature $
\bar{\tensor{
\kappa}}$ via in-thickness variations, reminiscent of the bending of a bi-metal strip \cite{timoshenko1925analysis, van2017growth}. However, given they are controlled by independent mechanisms, it is possible for  $
\bar{
\tensor{\kappa}}$ and $
\bar{\tensor{a}}$ to not satisfy the compatibility conditions, meaning that there is no surface that achieves all the preferred characteristics. Such incompatibility between  $
\bar{
\tensor{\kappa}}$ and $
\bar{\tensor{a}}$ is the appropriate dimensionally reduced  version of general geometric incompatibility in bulk elasticity \cite{Efrati2009} and results in the physical shell always being internally stressed, even in their energy minimizing state.  These internal stresses in turn underpin the instabilities observed in active shells. A key aspect of the mechanics of thin shells is that deforming the metric from $\bar{\tensor{a}}$ requires stretching, with an energetic penalty proportional to thickness $t$, while deforming the extrinsic curvature from $
\bar{
\tensor{\kappa}}$ only incurs a much smaller bending energy cost proportional to $t^3$. Thus in thin shells, isometric bending deformations are favoured wherever possible.

A simple example of incompatibility and instability arises in a spherical bi-layer gel cap. If we assume the cap was prepared by casting directly in the spherical state, it will have two equal principal curvatures,  $1/R$, and homogeneous positive Gauss curvature $K=1/R^2$. These curvatures will also be the preferred values, meaning the cap is compatible and unstressed. The hallmark of this compatibility is that, if we dissect a thin strip from the cap, it will not relax in shape, but retain the same curvature it had whilst part of the cap. However, if we now imagine the upper layer swells mildly (or thermally expands) relative to the lower layer, this will increase the preferred extrinsic curvature beyond $1/R$ whilst having a negligible effect on the metric. Unfortunately, the shell cannot adopt this increased curvature without also increasing the Gauss curvature, which would require an energetically prohibitive stretch. Instead,  it will remain in the same shape, but now with internal stresses; if we now dissect out a thin strip, it will be released from its metric constraints and relax to its preferred curvature, as illustrated in Fig.\ \ref{fig:intro}b). In this case, further swelling of the top layer  ultimately triggers an instability in which the spherical cap looses rotational symmetry and ultimately folds \cite{Pezzulla2018}, as shown in Fig.\
\ref{fig:intro}b). Conversely, large swelling of the bottom layer promotes bends with opposite sign to the original state, as shown in Fig.\ \ref{fig:intro}c). These will eventually drive the snap inversion of the cap, reminiscent of  the inversion of a bimetallic dome used in kettle switches, and the snap of a rubber-popper toy \cite{pandey2014dynamics}. Thus, even in this simple case, we may see two different instabilities, driven by different senses of incompatibility between extrinsic and intrinsic preferred curvature.

Importantly, different instabilities involve different mechanics, as some may proceed purely isometrically, while others require stretching. This difference is reflected in the thickness dependence of the buckling threshold. For example, the folding of the spherical cap subject to a curvature load proceeds via isometries, so the stretching energy is zero, and the threshold is thus thickness independent, requiring only a mild excess curvature $\delta\bk=\kappa-\bk \sim 1/R$. Conversely, the inside-out snapping of the cap inevitably requires stretching and is thus non-isometric. In reality, this instability nucleates at the boundary of the system where bend and stretch compete in a layer of width $\sim1/\sqrt{Rt}$, leading to a much higher buckling threshold that scales like $\delta\bar{\kappa} \sim 1/\sqrt{Rt}$ \cite{Pezzulla2018}. Finally, a fully closed spherical shells has no isometries available and there is no boundary,  leading to a yet-higher buckling threshold scaling like $\delta\bar{\kappa} \sim 1/t$ \cite{Pezzulla2018}.

By combining casting with initial preferred curvature, gel-like swelling spontaneous shape changes, and nematic-elastomer/muscle-like uniaxial spontaneous shape changes, a shell may undergo almost arbitrary programmed changes to $\tensor{\bar{a}}$ and $\tensor{\bar{\kappa}}$ \cite{ware2015voxelated,aharoni2018universal,barnes2019direct, van2017growth}. In this article, we offer a simple and comprehensive treatment of curvature driven instabilities in thin, shallow shells with arbitrary but homogeneous preferred curvatures. In such very-thin shells, the different scaling of stretching and bending means that the stretching energy turns into a constraint on the metric and hence the Gauss curvature, so one only has instabilities that proceed isometrically, with mild threshold curvature loads $\delta \bar{\kappa} \sim \sqrt{K}$. Nevertheless, despite this limitation, the model includes a wide variety of instabilities, in which the principle curvatures of the shell \textit{break-symmetry}, \textit{rotate} and \textit{invert}. The first type of instability is associated with supercritical spontaneous symmetry breaking and comes in two flavours: the continuous symmetry breaking of isotropy, and discrete breaking of up/down symmetry. The \textit{rotation} instability is a 90 degree rotation between states of broken isotropy which occurs via a Goldstone-like mode, characterised by a change in configuration but not in energy. Conversely, the \textit{inversion} instability occur as a subcritical jump between states of broken up/down symmetry and are associated with bi-stability. 

We summarise these results by plotting phase diagrams for active shells with zero, positive and negative Gauss curvature, which show how the state of the shell changes with preferred curvature, and where the different instabilities arise. Importantly, our model uncovers new instabilities, especially in the previously little discussed negative Gauss curvature systems. It also offers a much simpler perspective on several known instabilities which had previously been understood using more-complicated boundary layer physics and/or ad-hoc assumptions. Finally, although our model is for shallow shells, in the final section, we show that the threshold it predicts for the folding of a shallow spherical cap also extends to deep spherical caps.


\section{Geometric incompatibility and the isometric energy}\label{sec2a}
Our task is to find the achieved shape of a shallow and thin  shell with a homogeneous preferred curvature $\bar{\tensor{\kappa}}$, and also a preferred metric $\bar{\tensor{a}}$ that, via the theoroma egregium, encodes a homogeneous Gauss curvature $K$. Since the shell is thin, we may assume it adopts a final form that exactly achieves $\bar{\tensor{a}}$, and hence bears no stretch energy. Furthermore, since the shell is shallow, the achieved surface is approximated by a quadratic form $r(x,y)=(x,y,\half \kappa_1 x^2+\half \kappa_2 y^2),$ where $\kappa_1$ and $\kappa_2$ are the achieved principal curvatures, and $(x,y)$ are Cartesian coordinates in a tangent plane at the origin, aligned with the  curvature tensor $\tensor{\kappa}=\text{diag}(\kappa_1, \kappa_2)$. If the preferred curvature is $\tensor{\bar{\kappa}}$, then the bending energy of the shell will be \cite{Efrati2009}:
\begin{equation}
\label{bending_energy}
\mathcal{E}_{b}=\frac{Et^3}{24(1+\nu^2)}\int \left[(1-\nu)\Tr{\tensor{\kappa}-\bar{\tensor{\kappa}}}^2+ \nu \left(\Tr{\tensor{\kappa}-\bar{\tensor{\kappa}}}\right)^2 \right]\, dA.
\end{equation}
Since the shell bears no stretch energy, the curvature, and thus shape achieved by the shell, corresponds to a minimum of this bend energy. However, the minimisation is constrained by the theorema egregium, which dictates that the  Gauss curvature $\kappa_1 \kappa_2$ of the surface is an intrinsic property of the metric, and hence must match $K$ as encoded by $\tensor{\bar{a}}$, giving a compatibility constraint between curvature and metric:
\begin{equation}
\label{Gauss}
K=\det{\tensor{\kappa}}=\kappa_1\kappa_2.
\end{equation}
Formally, there are two further compatibility equations, known as the Codazzi-Mainardi equations, which also feature the metric connection computed from $\tensor{\bar{a}}$. However, the Cartesian coordinate system ($x,y$) forms a normal coordinate system for $\tensor{\bar{a}}$, meaning it is not only orthonormal, but all the connections vanish at the origin, so the Codazzi-Mainardi are automatically satisfied near the origin (and hence throughout our shallow shell) provided $\tensor{\kappa}$ too is homogeneous, as it is in our quadratic surface.  Furthermore, since both $\tensor{\kappa}$ and $\bar{\tensor{\kappa}}$ are homogeneous, we can trivially evaluate the  integral in  \ref{bending_energy} to get
\begin{equation}
\label{bending_energy2}
\mathcal{E}_{b}=\frac{A E t^3}{24(1+\nu^2)}\left[(1-\nu)\Tr{\tensor{\kappa}-\bar{\tensor{\kappa}}}^2+ \nu \left(\Tr{\tensor{\kappa}-\bar{\tensor{\kappa}}}\right)^2 \right]
\end{equation}
where $A$ is the area of the shell. Therefore, our task is simply to minimise Eq.\ \ref{bending_energy2} over $\tensor{\kappa}$, subject to Eq.\ \ref{Gauss}. Before we proceed, it is convenient to further simplify the problem by introducing
\[
H=\half \Tr{\tensor{\kappa}}=\half(\kappa_1+\kappa_2)\,\,\,\,\,\,\,\,\text{and} \,\,\,\,\,\,\,\,\, D=\half(\kappa_1-\kappa_2),
\]
which represent the mean achieved curvature and mean difference in achieved curvature of the shell, and the corresponding quantities $\bH=\half (\bk_1+\bk_2)$ and $\bD=\half (\bk_1-\bk_2)$ for the preferred curvature. Substituting these into eqn.\ \ref{bending_energy2} and re-scaling to eliminate the prefactor, we see the problem is equivalent to minimising 
\begin{equation}
\label{bendsimple}
\tilde{\mathcal{E}_b}=(H-\bH)^2+\gamma (D-\bD)^2 + 4 D \bD \sin^2 \theta,
\end{equation}
where $\theta$ is the angle between the principal directions of $\tensor{\kappa}$ and $\bar{\tensor{\kappa}}$, and $\gamma=(1+\nu)/(1-\nu) \geq 1 $  captures Poisson effects. Minimisation over  $\tensor{\kappa}$ is now represented by minimisation over $H$, $D$ and $\theta$, while $\bD$ and $\bH$,  (preferred curvature) are held constant. Minimisation is subject to the reformulated Gauss constraint:
\begin{equation}
\label{Ksimple}
K=H^2-D^2.
\end{equation}

Before minimising the energy, we clarify two  subtleties in the new parameterization. First, a rotation of the achieved curvature by $\theta=\pi/2$ is equivalent to inverting the roles of $\kappa_1$ and $\kappa_2$, which is also achieved by sending $D \to - D$ without changing $H$ and $\theta$. Although this might appear as a redundancy, it is helpful to maintain both $D$ and $\theta$  when studying the stability of an equilibrium. Secondly, flipping the signs of  $D$, $\bD$, $H$ and $\bH$ leaves the energy unchanged. This is a consequence of the arbitrary definition of positive vs negative curvature. Thus, when studying instabilities, we can often concentrate on a restricted set of values of $\bD$ and $\bH$, then use these facts to extend our theory to all values of the preferred bend.

In the remainder of this section, we shall discuss this minimisation explicitly for the cases when $K=0$, $K>0$ and $K<0$. In each case, we shall present a 2D phase diagram, showing the achieved states of the shell as a function of $\bH$ and $\bD$. Active changes in the shell are then represented by changes in  the preferred curvatures, $\bH$ and $\bD$, causing the shell to trace out a path on the phase diagram, and instabilities occur when this line passes from one region to another, indicating a qualitative change in the shape of the shell. 


\section{Instabilities in shallow shells}\label{sec2}

\subsection{Gauss flat shells (cylinders)}\label{subsec2}

We start with the simple case of a flat system that is isometric to the plane, $K=\kappa_1 \kappa_2=0$. The behaviour of such shells is well studied \cite{mansfield1989bending,seffen1999deployment,kebadze2004bistable,Pezzulla2016,Pezzulla2017,Jiang2018}, but we revisit these results in the context of our simple model to clarify their origin, and set the context for Gauss-curved systems.  The key feature of a Gauss flat sheets is that the Gauss constraint $K=0$ requires at least one curvature to vanish. Without loss of generality, we take the vanishing curvature to be $\kappa_2$, i.e. ($\kappa_1, \kappa_2)=(\kappa_0,0)$ meaning $D=H=\half \kappa_0$, and the resultant surface is a cylinder. Inserting this Gauss constraint into the bending energy, we find
\[
\tilde{\mathcal{E}_b}=\gamma (H-\bH)^2+ (H-\bD)^2 +4  H \bD \sin^2 \theta,
\]
where we have chosen to use $H$ rather than $\kappa_0$ to describe the magnitude of the curvature for consistency with our later treatment of Gauss curved systems. Contour plots of the energy as a function of $H$ and $\theta$ are shown in Fig.\
 \ref{fig:K0phase}, for a range of different preferred curvatures (i.e.
 $\bH$ and $\bD$). We see plots with one or two minima, or even whole lines of minima, which correspond to the stable states of the shell at each preferred curvature. These minima are given by:
\begin{align}
\label{eq01}
\frac{\partial \tilde{\mathcal{E}_b}}{\partial \theta}&=4 H \bD \sin 2 \theta=0,\\
\label{eq02}
\frac{\partial \tilde{\mathcal{E}_b}}{\partial H}&=2(H-\bD+2\bD \sin^2 \theta +\gamma(H-\bH))=0.
\end{align}
 We can also evaluate also the hessian matrix of second derivatives which allows us to assess stability of solutions:
\begin{equation}
\mathbf{H}_{\tilde{\mathcal{E}_b}}=\left(
\begin{array}{cc}
 2 (1+\gamma) & 4 \bar{D} \sin 2 \theta  \\
 4 \bar{D} \sin 2 \theta & 8 H \bar{D} \cos 2 \theta  \\
\end{array}
\right).
\label{hess0}
\end{equation}
A given equilibrium solution is either a minimum, a saddle or a maximum if the eigenvalues of $\mathbf{H}_{\tilde{\mathcal{E}_b}}$ are  positive, mixed, or negative respectively. 

\subsubsection{Isotropic preferred curvatures, $\bD=0$}
\begin{figure}[t!]
    \centering
 \includegraphics[width = \textwidth]{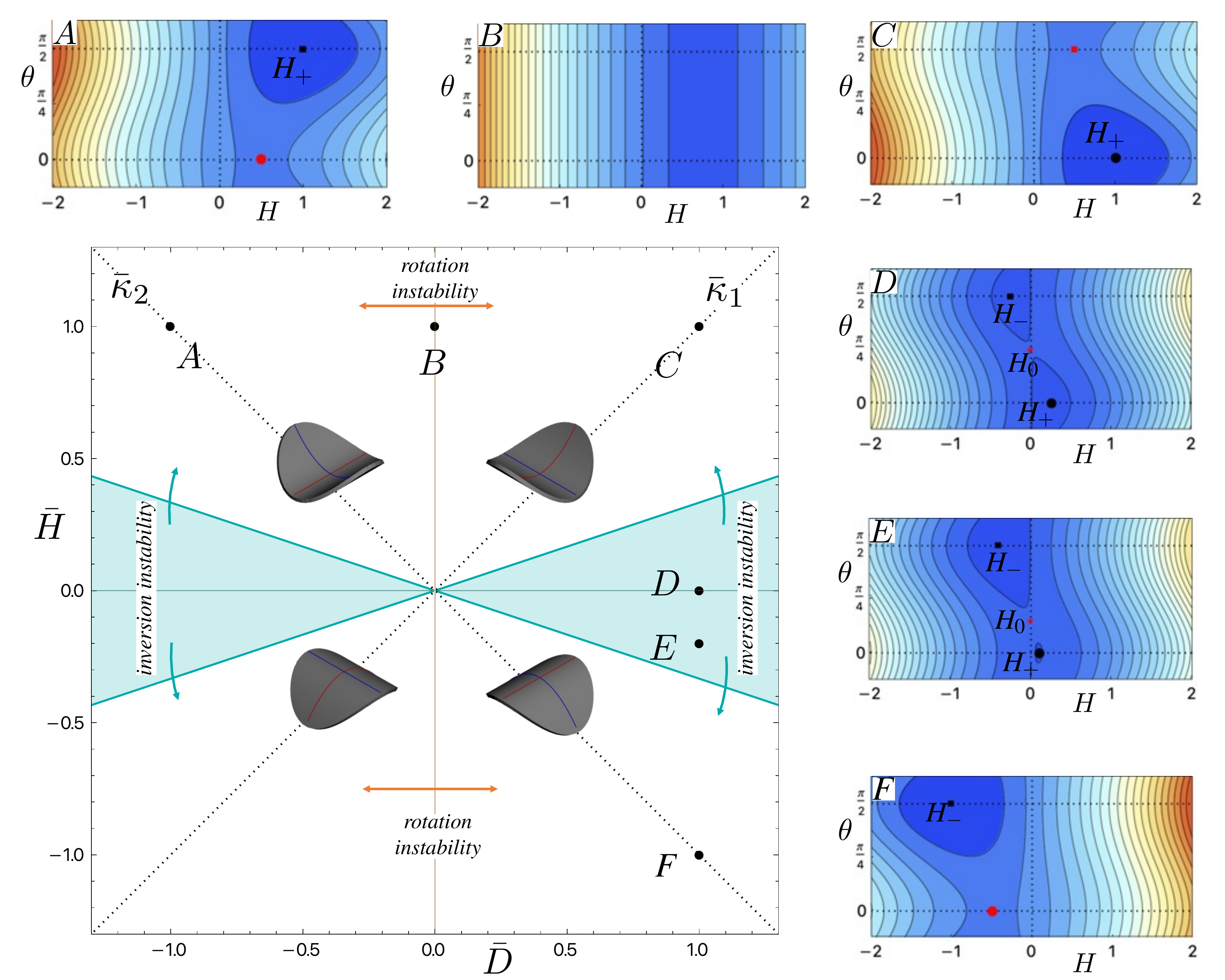}
    \caption{ The phase diagram for a Gauss flat system showing various states, accompanied by plots of their energy landscape, a bistable region (light blue shading) and the instabilities in the system.}
    \label{fig:K0phase}
\end{figure}

The simplest case arises if the preferred curvature is isotropic, meaning the preferred curvature is like a spherical cap  with $\bk_1=\bk_2$ and $\bD=0$. Such a system would arise in a bi-layer disk, in which the bottom layer swells or expands relative to the top \cite{mansfield1989bending, Pezzulla2016}. Of course, this preferred curvature violates the Gauss constraint $K=0$. However, this does not mean the plate remains flat, as the energy landscape forms a degenerate valley in the $\theta$ direction  at finite $H=\half \bH(1+\nu)$ \cite{mansfield1989bending} as seen in Fig.\ \ref{fig:K0phase}B.  Practically, this means the system chooses to roll-up like a cylinder to accommodate the preferred curvature. However, since the preferred curvature is isotropic, all rolling directions are equivalent, and by choosing one, the system breaks a continuous symmetry associated with in-plane isotropy. Within our isometric model, the plate will break symmetry and roll for  any finite $\bH$. In reality, a thick plates will actually initially stretch its mid surface and change $K$ to accommodate both preferred curvatures \cite{mansfield1989bending}, and these symmetry breaking isometric deformations occur past a threshold which scales like $t/R^2$ \cite{Pezzulla2016}, which indeed vanishes in the thin limit. 

\subsubsection{Equal and opposite preferred curvatures, $\bH=0$}
A second simple case arises if the preferred curvatures are equal and opposite, meaning $\bk_1=-\bk_2$ and $\bH=0$. Such a system would arise in a bilayer where the two layers contract uniaxially and equally, but in orthogonal directions, as seen in twisted nematic elastomer sheets \cite{sawa2011shape} or rubber bilayers with orthogonal pre-stretch \cite{efrati2014orientation, Jiang2018}. Again, this state of curvature is denied by the Gauss constraint.  As shown in Fig.\ \ref{fig:K0phase}D, the energy landscape now presents two equivalent minima at $(H,\theta)=(H_{+},0)$ and $(H_{-},\pi/2)$, separated by a saddle at $(0,\pi/4)$. Practically, this means the system will either roll upwards along the positive preferred curvature, or, equivalently, roll downwards along the negative one: the system breaks a discrete up-down symmetry,  the two configurations are energetically equivalent, and the system is bistable. As above, in the thin isometric limit, the plate will break symmetry and roll for  any finite $\bD$, though, in reality, there will be a small threshold that vanishes with thickness. This bistability has been studied in stiff  \cite{kebadze2004bistable} and soft \cite{Pezzulla2017, Jiang2018} sheets, and is used in cyclists snap-bands \cite{seffen1999deployment}. 

\subsubsection{General preferred curvatures}
At a general point on the phase diagram, away from these two special cases, there may be either one or two minimia. Near the $\bD=0$ line, the introduction of a small anisotropy in preferred curvature breaks the degeneracy of the $\theta$ valley, favouring one end over the other, so that the system is monostable, and rolls up into a cylinder along which ever direction has the largest preferred curvature, Fig.\ \ref{fig:K0phase}A,C. In contrast, near the $\bH=0$ line, the system remains bistable, Fig.\ \ref{fig:K0phase}E, but the symmetry between the two minima is lost, with the minima corresponding to the larger magnitude of preferred curvature being deeper. 

In this $K=0$ case, we may resolve this behaviour completely analytically. The angular minimisation condition,  equation \eqref{eq01}, is everywhere solved by $\theta=0$ and $\theta=\pi/2$, meaning the achieved and preferred curvature frames are aligned or anti-aligned. Substituting these into \eqref{eq02}, we find the corresponding $H$ values are $H=\half (1+\nu) \bH+\half (1-\nu)\bD$,  $H=\half (1+\nu) \bH-\half (1-\nu)\bD$, respectively, leading to the actual curvatures\cite{Jiang2018}
\begin{align}
\theta=0, \quad \kappa_0=\half(\bk_1+\nu \bk_2)\\
\theta=\pi/2, \quad \kappa_0=\half(\bk_2+\nu \bk_1).
\end{align}
These results show how an orthogonal preferred curvature modifies the achieved curvature via Poisson effects.
In the mono-stable region, one of these solutions is a saddle, and the other is the only minima, with the assignation delineated by the sign of $\bD$, to align $\kappa$ with the larger preferred curvature. We call the minimum state $H_+$ if $H>0$ and $H_-$  if $H<0$, as shown in the energy plots of Fig.\ \ref{fig:K0phase}. This allows us to easily  distinguish between rolled up or rolled down cylinders. Mathematically, we have
\begin{align}
\label{hplusminus}
H&=H_+ \equiv \half (1+\nu) \bH+\half (1-\nu)|\bD|\\
H&=H_- \equiv \half (1+\nu) \bH-\half (1-\nu)|\bD|.
\end{align}

In the bistable region there is a third solution to the Eq.\ \eqref{eq01}, $H=H_0=0$,which, via Eq.\ \eqref{eq02}, requires $\theta=\arctan \left(\sqrt{\frac{\bD+\gamma \bH}{\bD-\gamma \bH}} \right)$. This equilibrium state corresponds to the sheet being flat, and only exists in the bistable region $\bD \in [-\gamma \bH,\gamma \bH]$, shown in blue on the phase diagram, where both minima $H_{+}$ and $H_{-}$ coexist, and  $H_0$ forms a saddle between them. As one traverses the bistable region, the $H_0$ saddle moves between $H_{+}$ and $H_{-}$, with the collision and resulting merger giving the transition to mono-stability at the boundary.

\subsubsection{Phase diagram and instabilities}
The system thus has four basic configurations as shown as insets in the phase diagram, in which either the 1 ($\theta=0$) or 2 ($\theta=\pi/2$)  direction is curved, either up ($H_+$) or down ($H_-$). Each of these configurations is the global minimum in its  quadrant, and is given by:
\begin{center}
\begin{tabular}{ |p{2.5cm}|p{2.5cm}|p{2.5cm}|p{2.5cm}|}
 \hline
 
     K=0 & $\bD<0$ & $\bD>0$ \\ 
   \hline
        & $\theta=\pi/2$ & $\theta=0$ \\ 
   $\bH>0$  & $H=H_+$ & $H=H_+$ \\
       \hline
        & $\theta=0$ & $\theta=\pi/2$ \\ 
    $\bH<0$ &  $H=H_-$ & $H=H_-$ \\
   \hline
\end{tabular}
\end{center}

Here, intuitively, the sign of $\bH$ is the sign of the preferred mean curvature, where $H_+$ and $H_-$ are defined in \eqref{hplusminus}. Since the system can roll along one direction up or down, it will choose the curvature of the same sign as the preferred mean curvature. 
Importantly, each  minimum persists as a local minimum above/below its quadrant, within the bistable region where $|\bD|>\gamma |\bH|$.

Having found the states, we can identify two instabilities. Firstly, if a shell traverses the $\bD=0$ line, as shown by the sequence of states $A \to B \to C$, then its direction of achieved curvature will jump by 90 degrees when $\bD=0$. Physically, this occurs when a shell starts with a $\bar{\kappa}_1>\bar{\kappa}_2$, but then  $\bar{\kappa}_2$ grows, until it first equals, and then exceeds $\bar{\kappa}_1$. In response the shell rolls into a cylinder, first aligned along $\bar{\kappa}_1$, then, when the preferred curvature is isotropic $\bD=0$, all rolling directions are degenerate, then after, it rolls along $\bar{\kappa}_2$. The transition occurs as a rotation of 90 degrees between states of broken isotropy. We call such an instability a \emph{rotation instability}, as the frame of achieved curvature rotates. Curiously, as the rotation occurs along a degenerate valley, no energy is released, although the configuration of the shell does  change in a sub-critical way. This instability does not seem to have previously been discussed theoretically, but it was recently observed experimentally (using a bilyaer structure composed of a passive and singly-curved PET shell and a contractile nematic elastomer layer) and deployed for robotic locomotion \cite{gao2021molecularly}.

The second instability arises when we traverse through the bistable region, resulting in the shell inverting sub-critically by jumping between the two bistable states, as shown in the sequence $C \to D \to E \to F$. Physically, we start with a shell with $\bar{\kappa}_1>0$, $\bar{\kappa}_2=0$, so that the shell starts in a monostable region with a  single minima at $H_{+}$ (C), corresponding to rolling along $\bar{\kappa}_1$. If $\bar{\kappa}_2$ then becomes increasingly negative, the shell will enter the  bistable region, with a new local minima $H_{-}$ ( rolling along $\bar{\kappa}_2$), and a saddle between at $H_0$. However, the shell will remain stuck rolled along $\bar{\kappa_1}$, even as we pass the $\bH=0$ line, and the new minima becomes the global one. As the shell traverses the bistable region, the $H_{+}$ minima moves towards $H=0$ due to Poisson effects, and, finally, merges with $H_0=0$ to becomes a saddle, so the shell can jump, sub-critically, to  $H_{-}$ and roll negatively along $\bar{\kappa_2}$. This jump occurs between states of broken up/down symmetry and is thus an \textit{inversion instability}, that swaps both the direction and the sign of its achieved curvature. The threshold for instability is 
\begin{align}
\bH = -\bD/\gamma  \quad \iff  \quad  \bar{\kappa}_2=-\bar{\kappa}_1/\nu \label{K0inversion_threshold},
\end{align}
and the instability is a true sub-critical instability in which the configuration of the shell jumps, and a finite amount of energy is released.

The inversion instability of $K=0$ systems has been studied in \cite{Pezzulla2017} and \cite{Jiang2018}, but with results slightly different to those presented here. In the first case, an isometric model was considered, as here, but the instability threshold was established by equating the energies of the two configurations, rather than at the limit of bi-stability.  Conversely, in the latter work, the instability was addressed by studying the boundary layer in which curvature and stretch compete, leading to a thickness dependent threshold: this is surely important in thick systems, but the isometric model should suffice for thin systems, and clarifies that  the key physics of inversion arises isometrically in the bulk of the plate, rather than at its boundary. 

Finally, we remark the scale invariance of the Gauss-flat system. By assuming the system is thin and Gauss flat, we have removed any length scale from the problem. Thus, the only relevant piece of information needed to predict the behaviour of the shell is the relative magnitude of $\bH$ and $\bD$, while their absolute magnitudes is irrelevant, leading to the phase-diagram being delineated by straight lines.  

\subsection{Positive Gauss Curvature}\label{posgauss}
Let us now discuss shells with positive Gaussian curvature in which the two principal curvatures must have the same sign to satisfy $K=\kappa_1 \kappa_2>0$. We now chose to solve the Gauss constraint in eqn. \eqref{Ksimple} for $H$ and obtain $H=\pm \sqrt{K+D^2}$, so that the resultant energy is now a function of $D$, which is the natural order parameter for the cap folding instability highlighted in the introduction. This choice of the sign of $H$ corresponds to the shell breaking the discrete up-down symmetry of the flat state as a consequence of its metric: all $K>0$ shells must break this symmetry. The choice of sign describes two different set of states. The positive sign corresponds to the set of positive curvatures, where $\kappa_1, \kappa_2 >0$, and therefore $H>\sqrt{K}$, while the negative sign corresponds to a set of inverted states with $\kappa_1, \kappa_2<0$, $H<-\sqrt{K}$. Importantly, unlike in the flat case, there are no isometric paths between these two sets, as any such path would have to pass a point where at least one curvature vanished, giving $K=0$ and violating the  Gauss constraint. Thus, the two sets are disconnected and can be treated separately.

From the bending energy of the shell
\[
\tilde{\mathcal{E}_b}=\gamma(\pm\sqrt{K+D^2}-\bH)^2+ (D-\bD)^2 +4  D \bD \sin^2 \theta,
\]
we note that changing the sign of $\sqrt{K+D^2}$ (and therefore choosing the other set of states), or changing the sign of $\bH$ both lead to the same energy. This means that results for one set of states can be mapped to the other set by simply flipping the signs of both $H$ and $\bH$. 
We therefore fix $K=1$, equivalent to rescaling every curvature by $\sqrt{K}$, and consider the positive set of states with $H=\sqrt{D^2+1}>0$.  Contour plots of the energy as a function of $D$ and $\theta$ are shown in Fig.\
 \ref{fig:K1phase}, for a range of different preferred curvatures, again showing states with different patterns of minima. Importantly, the physical state of the cap is unchanged if we flip $D \to -D$ and send $\theta\to \theta\pm \pi/2$, so each physical state of the cap appears twice on each plot, once with $D$ positive, and once with negative. The equilibrium states can be found by minimising this energy w.r.t. variations in $\theta$ and $D$, leading to:
\begin{align}
\label{eq11}
 D \bD \sin 2 \theta=&0,\\
\label{eq13}
D \left(\frac{1+\gamma}{\bH \gamma}\right)-\frac{\bD}{\bH \gamma} \cos 2\theta=&\frac{D}{\sqrt{1+D^2}}.
\end{align}

We may also compute the Hessiam matrix of second derivatives to characterise the stability of solutions:
\begin{equation}
\mathbf{H}_{\tilde{\mathcal{E}_b}}=\left(
\begin{array}{cc}
 2+2 \gamma \left(1-\frac{\bH}{(1+D^2)^{3/2}}\right)& 4 \bar{D} \sin 2 \theta  \\
 4 \bar{D} \sin 2 \theta & 8 D \bar{D} \cos 2 \theta  \\
\end{array}
\right).
\label{hessp}
\end{equation}

\subsubsection{Isotropic preferred curvature, $\bD=0$}\label{foldingcap}
\begin{figure}[t]
    \centering
 \includegraphics[width = \textwidth]{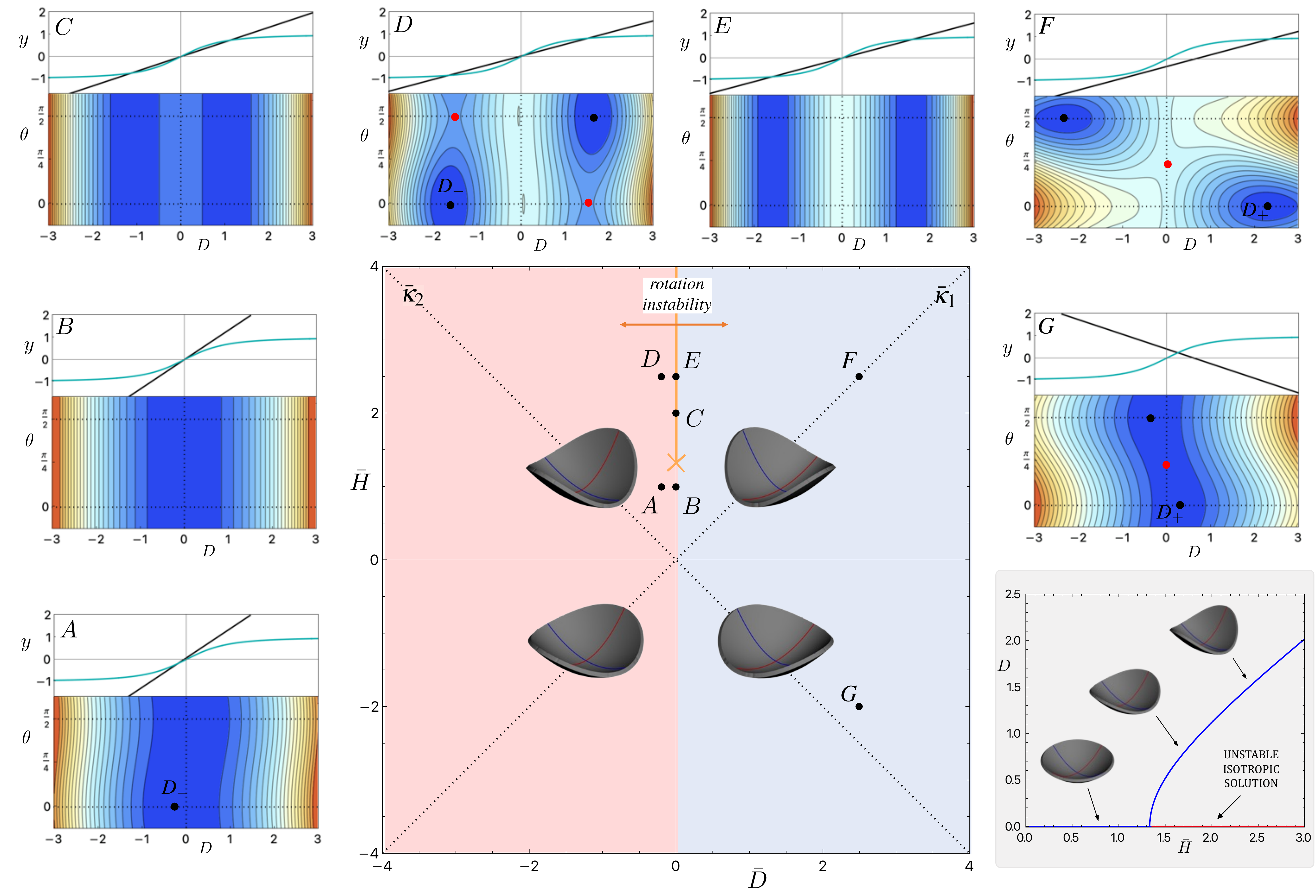}
    \caption{Phase diagram and energy contours plots of various states for a system with positive Gauss curvature. Above each energy plot, we show the intersection between the two lines in equation \eqref{eq13C}. On the bottom right, a plot of the amplitude of the symmetry breaking instability as a function of $\bH$. }
    \label{fig:K1phase}
\end{figure}

Again, the simplest case arises when the preferred curvature is isotropic, $\bk_1=\bk_2=\bH$ and $\bD=0$. Physically, this is the case of the bi-layer spherical cap, in which one layer swells relative to the other \cite{Pezzulla2018}. The isotropic preferred curvature is isotropic meaning any $\theta$ dependence in the energy is lost, $\tilde{\mathcal{E}_b}=\gamma(\sqrt{1+D^2}-\bH)^2+ D^2$. Naively, one might think the system chooses an isotropic state ($D=0$) to maintain the rotational symmetry. From state $B$ in Fig.\
\ref{fig:K1phase}, we note this is true when the preferred isotropic curvature $\bH$ is smaller than a critical value given by 
\begin{equation}
\bH_c=\frac{2}{1+\nu},
\end{equation}
leading to an energy landscape with a  degenerate minimum valley at $D=0$, corresponding to the rotationally symmetric isotropic state. We call this state the $D_0$ state. However, as shown in state $C$ in Fig.\ \ref{fig:K1phase}, when the isotropic preferred curvature is large, $\bH>\bH_c$, $D_0$ becomes a maximum ridge, and two new minima with $D_{\pm}= \pm \half \sqrt{\bH^2 (1+\nu)^2-4}$ emerge, creating an energy landscape with two degenerate valleys at finite $D$. The two valleys each describe the same set of physical states, which are states in which the cap has broken symmetry and is folded, having chosen to accommodate the larger preferred curvature in one direction, at the cost of even less curvature in the orthogonal direction to maintain $K$. Since the preferred curvature is isotropic, the larger curvature may fold the shell in any direction. However, as in the flat case, by choosing a folding direction the shell breaks a continuous symmetry. This time the preferred curvature threshold is non-vanishing and arises when the preferred curvature is larger than that allowed by the Gauss constraint, meaning the transition can be observed by varying the curvature with a post-buckling scaling $|D| \sim \sqrt{\bH-\bH_c}$, as shown at the bottom right of Fig.\ \ref{fig:K1phase}. Interestingly, weaker and negative preferred curvatures do not lead to an analogous instability, and the shell conserves its shape.

A graphical perspective can help understand this instability. Since eqn. \eqref{eq11} is automatically satisfied when $\bD=0$, the details of the instability are all contained in eqn. \eqref{eq13}, which becomes,
\begin{equation}
D \left(\frac{1+\gamma}{\bH \gamma}\right)=\frac{D}{\sqrt{1+D^2}},
\label{eq13B}
\end{equation}
and must be solved for $D$. The function on the LHS is a straight line with gradient $\frac{1+\gamma}{\bH \gamma}$ while on the RHS we have a sigmoidal function which smoothly transitions from $-1$ to $1$, and solutions are intercepts between these two graphs. Examples are shown above the energy plots in Fig.\ \ref{fig:K1phase}. Increasing $\bar{H}$ makes the gradient of the straight line shallower, causing a transition from one intersection at $D=0$ (state B) to three intersections at $D=D_-,0,D_+$ above the critical value $H_c$, (states C and E).

This instability has been observed experimentally in spherical caps made of bi-layer swelling gels \cite{Pezzulla2018}. However, the loss of symmetry was attributed to the competition between bend and stretch at the boundary layer. Our simple model shows that the same instability is predicted by isometric deformations of the bulk, again offering a simpler perspective on the  phenomena. 

\subsubsection{General preferred curvature}
If the preferred curvature is anisotropic, $\bD \neq 0$, the symmetry of the preferred curvatures is broken, meaning one principal preferred curvatures is larger than the other. In this case, eqn.\ \ref{eq11} is solved by $\theta=0, \pi/2$, indicating the achieved curvature will align or anti-align with the preferred curvatures. Indeed, if we start from the  isotropic case with $\bH<\bH_c$ and $\bD=0$ (state B) and then introduce a small asymmetry in the preferred curvature by increasing $\bD$, the result is to both skew the degenerate valley and break its degeneracy (state A) so that it has (physically equivalent) minima at $\theta=0,\pi/2$ separated by a saddle at $D=0$, $\theta=\pi/4$. The system is thus monostable, with an asymmetric achieved curvature that mirrors the preferred one. For the $\theta=0$ state, \eqref{eq13} becomes:
\begin{align}
\label{eq13C}
D \left(\frac{1+\gamma}{\bH \gamma}\right)-\frac{\bD}{\bH \gamma}=\frac{D}{\sqrt{1+D^2}}, 
\end{align}
and the solution gives the minimising value of $D$.  In analogy with the flat case, we call the  $D_-$ for a minimum of the energy with $D<0$ and $D_+$ for a minimum with $D>0$. Although this equation does not admit analytic solutions,  we see that this is still the intersection of a line and a sigmoidal function, and the effect of $\bD$ is simply a vertical offset in the line, moving the solution away from the origin.

Conversely, if we start from an isotropic preferred curvature  with $\bH>\bH_c$, we start with a landscape with two degenerate valleys (state C), describing folded-cap states, where the achieved curvature is already asymmetric, but may align in any direction. Now the introduction of asymtery in prefered curvature breaks the degeneracy of the valley, so that it has a minima at one end, corrosponding to the fold being aligned with the greater preferred curvature, and a saddle at the other, with anti-alignment (state D). The two valleys still describe the same set of physical states, so the two minima in plot D are the same physical state, and the shell is still mono-stable. Again, the minimising $D$ is not available analytically, but we may find the $\theta=0$ solution for $D$ graphically by considering eqn.\ \eqref{eq13C}:  the straight line is offset vertically, but there are still three intersections corresponding to a minima (left) maxima (middle) and saddle (right), as shown in plot D. The physically equivalent solutions for $\theta=\pi/2$ could be obtained from an analogous graphical method.

As we further increase $\bD$, and hence the vertical offset of the line, we will ultimately transition from having three intersections to having one, recovering a landscape like state F, with two minima and a saddle. The transition between these two types of landscapes occurs at $\bD^2= \frac{\left(\sqrt[3]{2(\nu +1) \bH^2}-2\right)^3}{2 (1-\nu)^2}$. Prior to this transition, a 1D cut of the the energy along the $\theta=0$ line appears double-welled, and the transition marks a transition to this cut being single welled. However, one must not conclude that this is therefore a transition from bistable to monostable, as, in every case, the subsidiary minima in the $D$ direction is unstable in the angular direction, and hence a saddle overall. In reality, the system is monostable for every finite $\bD$.

\subsubsection{Phase diagram and instabilities}
Bringing this all together, we may summarise the stable states of the spherical cap as
\begin{center}
\begin{tabular}{ |p{1cm}|p{2.5cm}|p{2.5cm}|}
 \hline
    K=1 & \bD<0 &\bD>0 \\ 
   \hline 
   & $D=D_+(\bH)$& $D=D_-(\bH)$  \\
   $H>0$         & $H=\sqrt{1+D^2}$ & $H=\sqrt{1+D^2}$ \\ 
   & $\theta=0$ & $\theta=0$\\
   \hline
\end{tabular}
\\
\begin{tabular}{ |p{1cm}|p{2.5cm}|p{2.5cm}|}
 \hline
    K=1 & \bD<0 &\bD>0 \\ 
   \hline 
   & $D=D_+(-\bH)$& $D=D_-(-\bH)$  \\
   $H<0$         & $H=-\sqrt{1+D^2}$ & $H=-\sqrt{1+D^2}$ \\ 
   & $\theta=0$ & $\theta=0$\\
   \hline
\end{tabular}
\end{center}
where the two tables show the two disconnected sets of states with $H>0$ (top) and $H<0$ (bottom) and we have not explicitly included $\bH$ in the table since changing its sign does not affect the nature of the solution. Importantly, $D_{-}(\bH)<0$ is the left-most root of eqn.\ \eqref{eq13C}, while $D_{+}(\bH)>0$ is the right-most root. The values of the minima of the inverted set of states can be thought of in two ways. The straightforward way is to consider the other sign of $H$, meaning $H<0$ is the inverted cap. However, this cap can be turned upside down, meaning $H>0$ once again, but now $\bH$ has changed sign. Thus, minima of the inverted states can be found by replacing $\bH$ with $-\bH$ in the equations for the original spherical cap, as shown in the table.

As illustrated in Fig.\ \ref{fig:K1phase}, across the $D=0$, the $D_+$ and $D_-$ states merge continuously via $D=0$ below $\bH_c$, but are discontinuous above $\bH_c$, where they are equal and opposite. This feature leads to two instabilities. First, as already discussed, if we have isotropic but increasing preferred curvature, $\bD=0$, $\bH>0$, then the shell remains isotropic for $H<H_c$, but breaks isotropy and folds in a \textit{continuous symmetry breaking} instability for $H>H_c$. This is the trajectory $B-C-E$ on the diagram, and leads to a super-critical growth of $D$ beyond $H_c$, as shown in the bottom right of Fig.\ \ref{fig:K1phase}. Secondly, if the shell crosses the $\bD=0$ line above $H_c$, $D\to E\to F$, it will jump discontinuously between states of broken isotropy, from $D_{-}$ to $D_{+}$. At the moment of discontinuity, the sign of $D$ simply flips, so this is a \textit{rotation instability}, in which direction of greater folding rotates by 90 degrees, so that it always matches the larger preferred curvature. As in the Gauss-Flat state, it produces a finite change in the configuration of the shell, but no release of energy. 

Our isometric model does not allow transitions between the two disconnected sets of states, which would be \textit{inversion instabilities}. In reality such instabilities do occur, but via non-isometric pathways, with the stability threshold which diverges for very thin shells and on the the depth of the cap \cite{Taffetani2018, Pezzulla2018}.

\subsection{Negative Gauss curved surfaces}
\begin{figure}[t!]
    \centering
 \includegraphics[width = \textwidth]{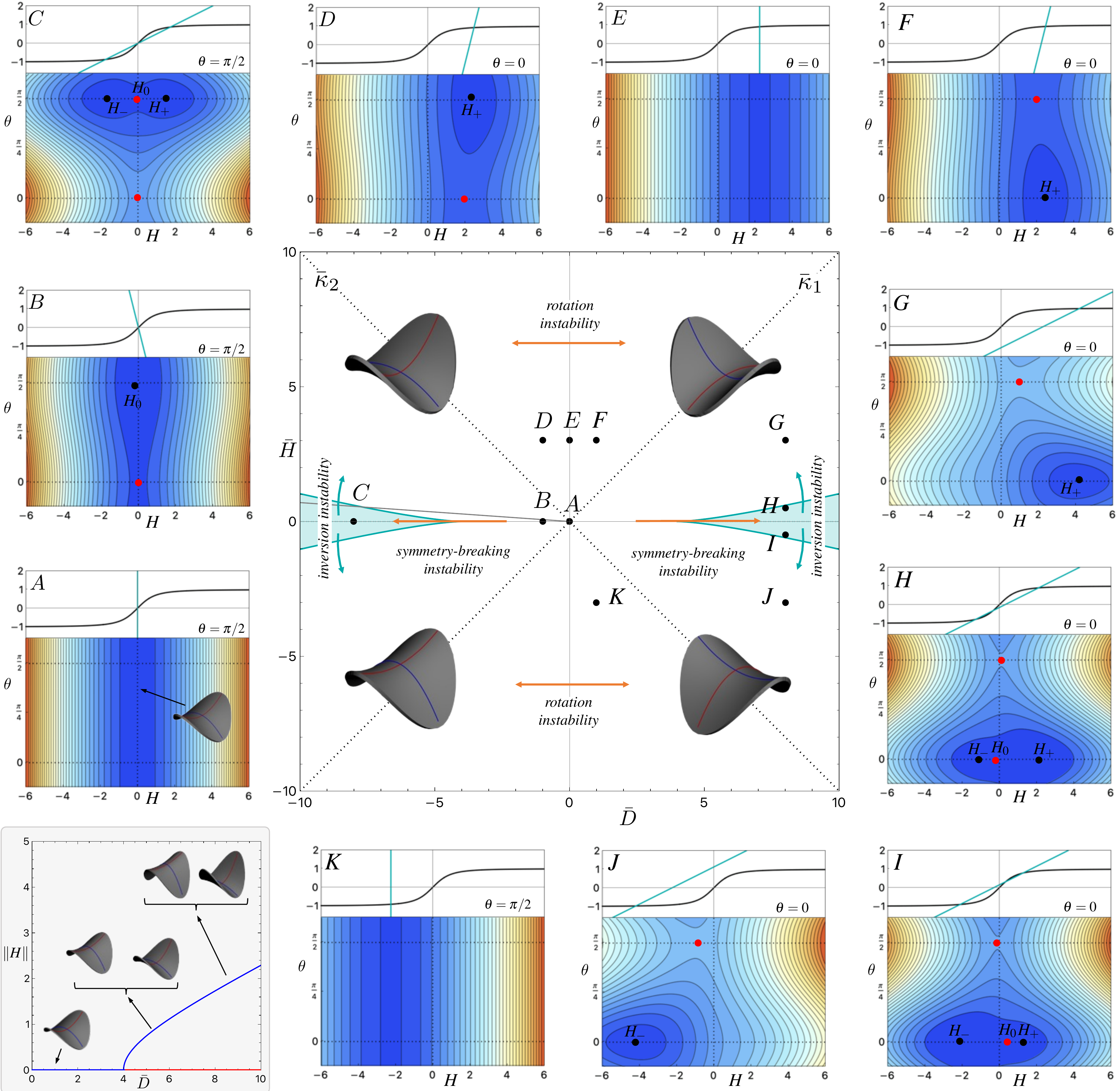}
    \caption{Phase diagram and energy contours plots of various states for a system with Negative Gauss curvature. Above each energy plot, we show the intersection between the two lines in equation \eqref{eqn12}. The bistable region is shaded in light blue. On the bottom left, a plot of the amplitude of the symmetry breaking instability as a function of $\bD$. }
    \label{fig:KNphase}
\end{figure}

Finally, we look at shells with negative Gaussian curvature, $\kappa_1 \kappa_2=K<0$, in which the two principal achieved curvatures must have opposite sign forming a saddle. The metric thus requires a choice of orientation of the principal direction, meaning that all $K<0$ shells break the isotropy of the flat state. To write down the bending energy, we first solve the Gauss constraint in equation \eqref{Ksimple}. As we have done in the flat case, we return to solving for $D$ obtaining $D=\pm \sqrt{K+H^2}$. Without loss of generality, we set $K=-1$ and pick the positive solution $D=\sqrt{1+H^2}$, leading to the bending energy:
\begin{equation}
\label{enn}
\tilde{\mathcal{E}_b} = \gamma(H-\bH)^2+ (\sqrt{H^2+1}-\bD)^2 +4  \sqrt{H^2+1} \bD \sin^2 \theta.
\end{equation}
Again, this energy is plotted for a range of preferd curvatures in Fig.\ \ref{fig:KNphase}. The equilibrium states of the system can be found by minimising this energy w.r.t. to variations in $\theta$ and $H$, leading to the equations:
\begin{align}
\label{eqn11}
\bD \sqrt{1+H^2} \sin 2 \theta=0,
\end{align}
\begin{align}
\label{eqn12}
\frac{H (1+\gamma)}{\bD \cos 2\theta}-\frac{\bH \gamma}{\bD \cos 2\theta}=\frac{H}{\sqrt{1+H^2}}.
\end{align}
The first equation is solved by  $\theta=0$ or $\theta=\pi/2$ --- indicating the bends are aligned or anti-aligned with the preferred ones --- or by $\bD=0$, indicating the preferred curvature is isotropic, so the energy is insensitive to orientation. The second equation contains much more complexity and, similar to the positive Gaussian curvature case, is composed of a straight line on the LHS and a sigmoidal function on the RHS. However, in this case the main variable is $H$ and the offset depends on $\bH$.

Finally, we also evaluate the hessian matrix that will allow us to discuss the stability of equilibrium solutions, 
\begin{equation}
\label{hessn}
\mathbf{H}_{\tilde{\mathcal{E}_b}}=\left(
\begin{array}{cc}
 2\left(1+\gamma-\frac{\bD \cos 2\theta}{(1+H^2)^{3/2}}\right)&  \frac{4 \bD H \sin 2 \theta}{\sqrt{1+H^2}}   \\
 \frac{4 \bD H \sin 2 \theta}{\sqrt{1+H^2}} & 8 \bD \sqrt{1+H^2} \cos 2 \theta  \\
\end{array}
\right).
\end{equation}

\subsection{Zero preferred curvatures, $\bH=\bD=0$}
We first consider a shell with zero (flat) preferred curvature, i.e. $\bD=\bH=0$. In this case, a positive or flat Gauss shell achieves a state which mirrors the preferred rotational symmetry by taking the isotropic spherical cap or fully flat state respectively. However, a negative Gauss surface must form a saddle and therefore cannot maintain the rotational symmetry. This key difference is what characterises much of the behaviour of negative surfaces. For the negative $K$ case, we obtain an  energy landscape of state  like $A$ in Fig.\ \ref{fig:KNphase}, with a degenerate valley at $H=0$ (and hence $D=1$) corresponding to a state in which the two principal curvatures are equal in magnitude ($\kappa_1=-\kappa_2=1$,  $H=0$) but in any orientation. We call such a state with equal magnitude curvatures  a \textit{symmetric} saddle. 

\subsection{Isotropic preferred curvatures, $\bD=0$}
We now turn to isotropic but non-zero preferred curvature, meaning $\bk_1=\bk_2$, $\bD=0$ and $\bH\neq 0$. An energy plot for $\bH>0$ is shown in state $E$ of Fig.\ \ref{fig:KNphase}, clearly highlighting a valley at $H=\half \bH (1+\nu)>0$ which, naturally, is degenerate in $\theta$. These states are asymmetric saddles, with the larger magnitude curvature having the same sign as the  preferred curvature. However, since the preferred curvature is isotropic, the achieved asymmetric saddle can be oriented along any direction.

\subsection{Symmetric preferred saddle, $\bH=0$}
There is another regime, important in  systems of negative Gaussian curvature, in which the preferred curvature itself is a symmetric saddle, meaning the two preferred curvatures have equal magnitudes but opposite sign, $\bk_1=-\bk_2$ ($\bH=0$ and $\bD \neq 0$) giving the energy a discrete up-down symmetry.  Naively, we may think this preferred curvature always favours a symmetric saddle ($H=0$) with matching curvatures, i.e. aligning the positive curvature with the preferred positive curvature and the negative with the preferred negative. As shown by the energy plot of state $B$ in figure \ref{fig:KNphase}, this is true when the preferred saddle is weak, since the energy landscape for $\bD<0$ forms a valley about $H=0$ with a minimum at $\theta=\pi/2$ and a saddle solution at $\theta=0$. The minimum and the saddle correspond to the achieved saddle being aligned and anti-aligned with the preferred one.

However, as shown in state $C$ of Fig. \ref{fig:KNphase}, when the curvature load $|\bD|$ is larger than a critical value 
\begin{equation}
    \bD_c=\frac{2}{1-\nu},
\end{equation}
the minimum in the valley splits into two symmetric minima at $H_{\pm}=\pm \half \sqrt{\bD^2(1-\nu)^2-4}$ and $\theta=\pi/2$. The two minima give rise to a bistable region, reflecting a preference of the system to partially accommodate either one of the two large preferred principal curvatures rather than remaining symmetric and accommodating neither. The same behaviour is observed when $\bD>0$ and $\bD>\bD_c$, with a minimum at $\theta=0$ instead. The choice of which curvature is accommodated is associated with the breaking of the energy's discrete up/down symmetry. As shown in the inset on the bottom left of figure \ref{fig:KNphase}, this behaviour is a supercritical, leading to a post buckling amplitude scaling, $|H| \sim \sqrt{\bD-\bD_c}$. 

Curiously, the mathematics of this discrete symmetry breaking transition with $K<0$ are very similar to the continuous  symmetry breaking for $K>0$. As in the positive case, it is again helpful to consider the situation graphically. Without loss of generality, we take $\bH=0$, $\bD<0$, like state $C$. Equation \eqref{eqn11} is solved by  $\theta=0$ and $\theta=\pi/2$. To find the minima, we first look at the correctly matched saddle, with $\theta=\pi/2$. Then, eqn. \eqref{eqn12} becomes,
\begin{equation}
\label{eqn12b}
-H \frac{(1+\gamma)}{\bD}=\frac{H}{\sqrt{1+H^2}}.
\end{equation}
This structure is indeed familiar from the positive Gaussian surface, eqn. \eqref{eq13B}, where the gradient of the line determines the number of solutions. When $|\bD|$ is small, we have only one solution which we call $H_0$ and corresponds to the symmetric saddle with $H=0$. When $|\bD|>\bD_c$, the line intersects the sigmoidal function two more times generating solutions $H_{\pm}=\pm \half \sqrt{\bD^2(1-\nu)^2-4}$. These two solutions are minima, while the $H_0$ solution becomes a maximum as can be checked from the hessian matrix \eqref{hessn}.

When $\bD<0$, the state with $\theta=0$ corresponds to a saddle rotated the wrong way around and with mismatched curvatures. In this case, the second equilibrium condition eqn \eqref{eqn12} becomes $H \frac{(1+\gamma)}{\bD}=\frac{H}{\sqrt{1+H^2}}$ and has only solution $H=0$, a symmetric saddle state with the curvatures aligned in the wrong way around. This state is obviously unstable, as can be easily checked from the hessian matrix; the system wants to rotate to align its bends correctly. When $\bD>0$, the $\theta=0$ solution becomes the minimum while the $\theta=\pi/2$ becomes the saddle.

\subsection{General preferred curvature}
We have discussed what happens if we move a little away from the isotropic preferred curvature $\bD=0$ in the phase diagram (state $E$) by introducing a small anisotropy to the prefered curvature.  In this case, the effect is to break the angular degeneracy, leading to a state like $F$ in which there is a single minima, corresponding to an asymmetric achieved saddle with a determined alignment. It is also helpful to consider stepping away from the symmetric preferred saddle states with $\bH=0$, by introducing a small asymmetry in the magnitudes of the preferred curvature. If we start from a monostable state like $B$ with $|D|<D_c$, with a single symmetric-saddle minima at $H=0$, then the small asymmetry simply moves this minima away from the origin, as in state $D$, leading to a monostable landscape in which the shell adopts a mildly asymmetric saddle reflecting the  now asymetric prefered curvatures. In contrast, if we start from a bistable situation, like $C$, with $|D|>D_c$, then asymmetry initially breaks the symmetry between the two minima, which become a global minima where the achieved asymmetry is aligned with the preferred, and a local minima where it is anti-aligned, as seen in state $H$. It follows that there is a region of bistablity on the phase diagram, and a region of monostability. 

We cannot find the values of $H$ for these general states analytically, but, again, it is helpful to consider a graphical approach. If we focus on $\bD>0$ (the right side of Fig.\ \ref{fig:KNphase}) the minima are found at $\theta=0$, which solve \eqref{eqn11}. The second equilibrium condition takes the form:
\begin{align}
\label{eqn13}
\frac{H (1+\gamma)}{\bD}-\frac{\bH \gamma}{\bD}=\frac{H}{\sqrt{1+H^2}},
\end{align}
which is once more the intersection of an offset line with the sigmoidal function. As seen in Fig.\ \ref{fig:KNphase}, the set of solutions are  exactly analogous to those discussed for the positive case, eqn. \eqref{eq13C}, with the system having only one solution when $\bD$ is small, but three solutions when $\bD$ is large. In the region with one solution, we call the minimum $H_{+}$ or $H_{-}$ depending on whether its $H$ value is greater or less than zero. Since there is only one minimum, the system is monostable. When there are three solutions, we classify the three roots, moving from left to right, as $H_{-}<0$, $H_0$ and $H_{+}>0$. The states  $H_{-}$ and $H_{+}$ are still minima which now coexist, and $H_0$ is a saddle, meaning the system is bistable. The global minima is the root with the same sign as $\bH$. An exactly analogous situation arises for $\bD<0$, but with $\theta=\pi/2$. 

At the limit of bistability, the local minima annihilates with the saddle, causing the top-right entry of the Hessian to vanish. This observation allows us to find the limit the bistable region as 
\begin{align}
\label{bistablen}
\bH^2= \frac{\left(\sqrt[3]{2(1-\nu)^2\bD^2}-2\right)^3}{2 (\nu
   +1)^2}.
\end{align}

\subsection{Phase diagram and instabilities}
We now can discuss all the (minimum energy) states in the system. Fist, we note that there are four main states, covering the combinations of the principal achieved curvatures signs and magnitude asymmetries. These four states are exemplified by the insets in the four quadrants of the phase diagram in Fig.\
\ref{fig:KNphase}, and are each the global minimum within their quadrant

\begin{center}
\begin{tabular}{ |p{1.cm}|p{2.5cm}|p{2.5cm}|p{2.5cm}|}
 \hline
 
 K=-1    & $\bD<0$ & $\bD>0$ \\ 
   \hline
           & $\theta=\pi/2$ & $\theta=0$ \\ 
    $\bH>0$ &  $H=H_+$ & $H=H_+$ \\
       & $D=\half \sqrt{1+H_+^2}$ & $D=\half \sqrt{1+H_+^2}$ \\
    \hline
        & $\theta=\pi/2$ & $\theta=0$ \\ 
   $\bH<0$  & $H=H_-$ & $H=H_-$ \\
       & $D=\half \sqrt{1+H_-^2}$ & $D=\half \sqrt{1+H_-^2}$ \\
   \hline
\end{tabular}
\end{center}
    where $H_{+}>0$ and $H_{-}<0$ are the right and left most roots of \eqref{eqn13}. Furthermore, each state persists as a local minimum beyond above/below its quadrant, within the region of bistability.  Importantly, $H_{+}$ and $H_{-}$ connect continuously via $H=0$ (a symmetric achieved saddle) across $\bH=0$ for $|\bD|<\bD_c$, but discontinuously for $|\bD|>\bD_c$, in the bistable region. 

The negative Gauss shell can thus sustain three types of instability. All $K<0$ saddles break the isotropy of a plane, even in the isotropic preferred case $\bD=0$.   Crossing the $\bD=0$ line,  $D\to E \to F$, thus leads to a \emph{rotation instability}, where the (asymmetric) achieved saddle will rotate through states of broken isotropy by 90 degrees, to conform with the preferred anisotropy. Conversely, a saddle does not by default break up-down symmetry, but moving along the $\bH=0$ line eventually triggers a super-critical \emph{symmetry breaking instability} at $\bD_c$, in which a symmetric saddle becomes a bistable asymmetric saddle, with one curvature larger in magnitude than the other, $B \to C$ and broken up/down symmetry.  Crossing through the resulting bistable region by varying $\bH$, path $G \to H \to I \to J$, will give a sub-critical \emph{inversion instability} where the shell jumps from $H_{+}$ to $H_{-}$ between states of broken up/down symmetry at the limit of bistability shown in eqn.\eqref{bistablen}.
During inversion, the saddle exchanges the asymmetry of its curvatures without exchanging their signs. 

\newpage

\section{Curvatures-driven loss of stability in deep spherical caps}

As discussed in section \ref{sec2}\ref{posgauss}\ref{foldingcap}, a shallow spherical cap subject to an excess preferred curvature looses rotational symmetry and folds when the preferred curvature exceeds a critical value $\bk_c=\frac{2}{1+\nu}\sqrt{K}$, promoting the folding of the structure. In this section, we show that critical buckling load and the 2-fold symmetric buckled shape predicted by the simple shallow model are correct even in deep shells, highlighting the shallow model's value.  

\begin{figure}[t]
    \centering
 \includegraphics[width = 0.8 \textwidth]{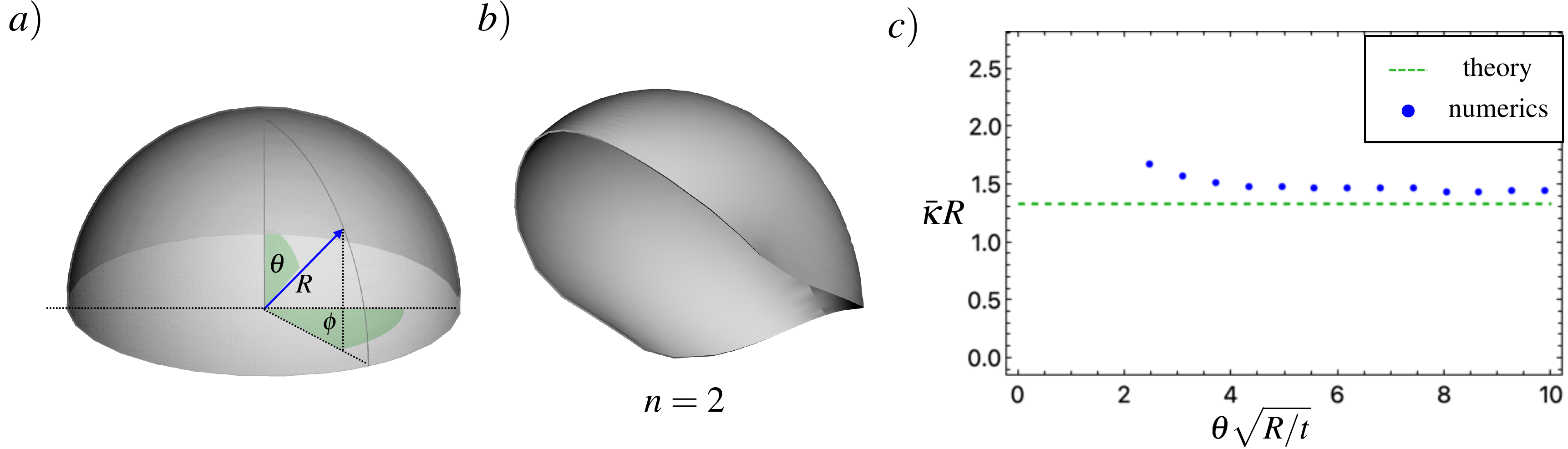}
    \caption{a) Schematics of the coordinate system used to describe a deep spherical cap. The cap has depth $\theta_0$.b) The folding deformation associated with the mode $n=2$ of the Rayleigh isometry. c) Comparison between numerical simulations and theoretical prediction for the buckling threshold inducing the loss of rotational symmetry in a spherical cap as a function of the cap depth $\bar{\theta}=\sqrt{R/t}\theta_0$. Numerics (blue points) for spherical caps of thickness $t/R=10^{-3}$ are taken from \cite{Pezzulla2018}, while our buckling threshold (dashed green line) is $\bk R=\frac{2}{1+\nu}.$ }
    \label{fig:deep}
\end{figure}

We start by considering a  spherical shell with angular depth $\theta_0$, and described with spherical coordinates $(R,\theta,\phi)$ as shown in Fig.\ \ref{fig:deep}a). We take an isotropic preferred curvature $\tensor{\bk}=\bk \tensor{I}$, as would result in a bilayer gel shell, and, without loss of generality, we set $K=1$. Following our work for the shallow shell, let us write the principal curvatures as:
\begin{align}
\kappa_1(\theta,\phi)&=H(\theta,\phi)+D(\theta,\phi)\\
\kappa_2(\theta,\phi)&=H(\theta,\phi)-D(\theta,\phi)
\end{align}
where $H=\half(\kappa_1+\kappa_2)>0$ is the mean curvature and $D=\half(\kappa_1-\kappa_2)$ is the curvature asymmetry. Note that $H$ and $D$ now depend on the position of the shell via $\theta$ and $\phi$. However, for sake of conciseness, we will avoid writing explicitly the dependence.  The Gauss constraint requires:
\begin{equation}
K=H^2-D^2=1
\end{equation}
which we satisfy by setting $H=\sqrt{1+D^2}$. 
This allows us to write the bending energy in the simple form:
\begin{equation}
\mathcal{E}_b=\frac{E t^3}{12(1-\nu^2)}\int \left[(\nu +1) \left(\bk-\sqrt{1+D^2}\right)^2- (\nu -1) D^2 \right] dA.
\label{b_energy}
\end{equation}
Note that such a simple form is only possible when the preferred bend is isotropic, as it does not require us to specify the orientation of the achieved curvature frame with respect to the preferred one.

Before instability the shell is spherical, so we have $D =0$, and we may conduct the integral to find the energy is $\mathcal{E}_b=\frac{E t^3}{12(1-\nu^2)}(\nu +1) \left(\bk-1\right)^2 A$ where $A$ is the area of the shell. If we now consider a small amplitude perturbation from this spherical state,  giving rise to a small $D(\theta,\phi)\neq 0$, the principal curvatures take the form:
\begin{align}
\kappa_1(\theta,\phi)&=1+D(\theta,\phi)+\half D(\theta,\phi)^2+...\\
\kappa_2(\theta,\phi)&=1-D(\theta,\phi)+\half D(\theta,\phi)^2+...
\end{align}
Similarly, inserting $D$ into the enery and expanding, we obtain:
\begin{equation}
\mathcal{E}_b=\frac{E t^3}{12(1-\nu^2)}\int \left[ (\nu +1) \left(\bk -1\right)^2+\left(2-(1+\nu) \bk\right) D^2+\frac{1}{4} (1+\nu) \bk D^4+...\right] dA.
\label{b_energy_b}
\end{equation}
Crucially, the coefficient of the second order term changes sign at the critical curvature load
\begin{equation}
\bk_c=\frac{2}{1+\nu}.
\end{equation}
Beyond this threshold, any perturbing function $D(\theta, \phi)$ will save energy compared to the spherical state, so the system has become unstable. Looking at the fourth order term, we see that beyond the threshold, we expect $|D|=\sqrt{\bk-\bk_C}$. Both these results make no assumptions about the depth of the shell, but exactly coincide with the those obtained earlier for shallow shells. 

Formally, the above argument is not quite sufficient to prove instability, as not every form of $D(\theta, \phi)$ corresponds to a real isometry of the surface. The issue arises because, although we have ensured the Gauss equation is satisfied, in deep shells, the curvatures must also satisfy the Codazzi Mainardi compatibility equations to ensure they describe a real surface. However, fortunately, a buckling isometry does indeed exist, as was identified by Lord Rayleigh in the study of vibration modes of shells \cite{strutt1877theory}. In the case of a sphere with a hole around the south pole, the first order isometry is given by:
\begin{align}
\delta R &=R \sum_n^\infty (n+\cos \theta )  \tan(\half \theta)^n \left(A_n\cos(n \phi)+ B_n \sin(n \phi)\right),\\
\delta \phi&=\sum_n^\infty \tan(\half \theta)^n \left(B_n\cos(n \phi)-A_n \sin(n \phi)\right),\\
\delta \theta&=-\sin\theta\sum_n^\infty   \tan(\half \theta)^n\left(A_n\cos(s \phi)+ B_n \sin(n \phi)\right).
\end{align}
where  $n$ is an integer which corresponds to the mode number of the solution, while $A_s$ and $B_s$ are amplitudes that may be fixed to any small value. We also note that modes with $n=0,\,1$ are solid body motion and rotation respectively, with no deformation. However, every mode with $n \geq 2$ gives $D\neq0$, and hence becomes unstable at our buckling threshold.  In Fig.\ 
\ref{fig:deep}b) we show the folding deformation induced by the mode $n=2$ isometry. A direct computation reveals that the $n=2$ mode indeed achieves the lowest energy of all the pure modes, in agreement with  numerically and experimental observations.

A detailed numerical and experimental study of this curvature induced instabilities in bi-layer spherical shells was recently presented by Pezzulla et. al. \cite{Pezzulla2018}. The authors use a combination of theoretical arguments and fitting to propose that the folding of a spherical cap with angular size $\theta_0$ occurs at the threshold curvature
$$\bar{\kappa} R =\frac{a}{\theta_0^2}\left(\frac{t}{R}\right)-c-\frac{b}{\pi-\theta_0}\sqrt{\frac{t}{R}}$$ 
where the parameter $a=\sqrt{10+7\sqrt{2}} $ was found by matching the theory to the flat plate case, while $b=3.6$ and $c=-0.98$ were found fitting experimental/numerical data with $\nu=0.5$. This result is more ambitious than the current work, as it aims to also capture the effect of boundary layer stretching, leading to a thickness dependent result, whereas we focus on the truly thin isometric limit. However, taking this limit of their result, in which $t/R$ vanishes, the predicted threshold would be $\bar{\kappa} =-c=0.98$, which contrasts with ours at  $\bar{\kappa} R= 4/3\approx 1.33$. In Fig.\ \ref{fig:deep}c, we compare our result with the thinnest data available in \cite{Pezzulla2018} ($t/R=10^{-3}$) for a range of angular depths of shell. We see an extremely encouraging level of agreement, particularly for the deeper shells. Agreement for shallower shells is actually somewhat worse because the shallow shells enter a regime where the boundary layer extent, $\sqrt{R t}$, covers a large fraction of the shell. We thus see that shallow theory actually applies in a delicate limit where the shell is geometrically shallow, but large and thin enough that the boundary layer is still negligible.

\section{Summary and Conclusions}

In this paper, we have studied a simple model of instabilities in active shells that arise from the geometric incompatibility of their intrinsic and extrinsic curvatures. We have proposed a simple model, based on the requirements that the shell: (1)  be geometrically shallow, (2) be thin enough to only deform isometrically, (3) has metric encoding homogeneous Gauss curvature $K$, and (4) is subject to thickness variation encoding homogeneous preferred curvature tensor $\tensor{\bar{\kappa}}$. If these conditions are met, the configuration of the shell is simply given by the minimisation of the bending energy over achieved homogeneous curvature $\tensor{\kappa}$, subject to the constraint that the achieved Gauss curvature is $K$.

\begin{figure}[t]
    \centering
 \includegraphics[width = \textwidth]{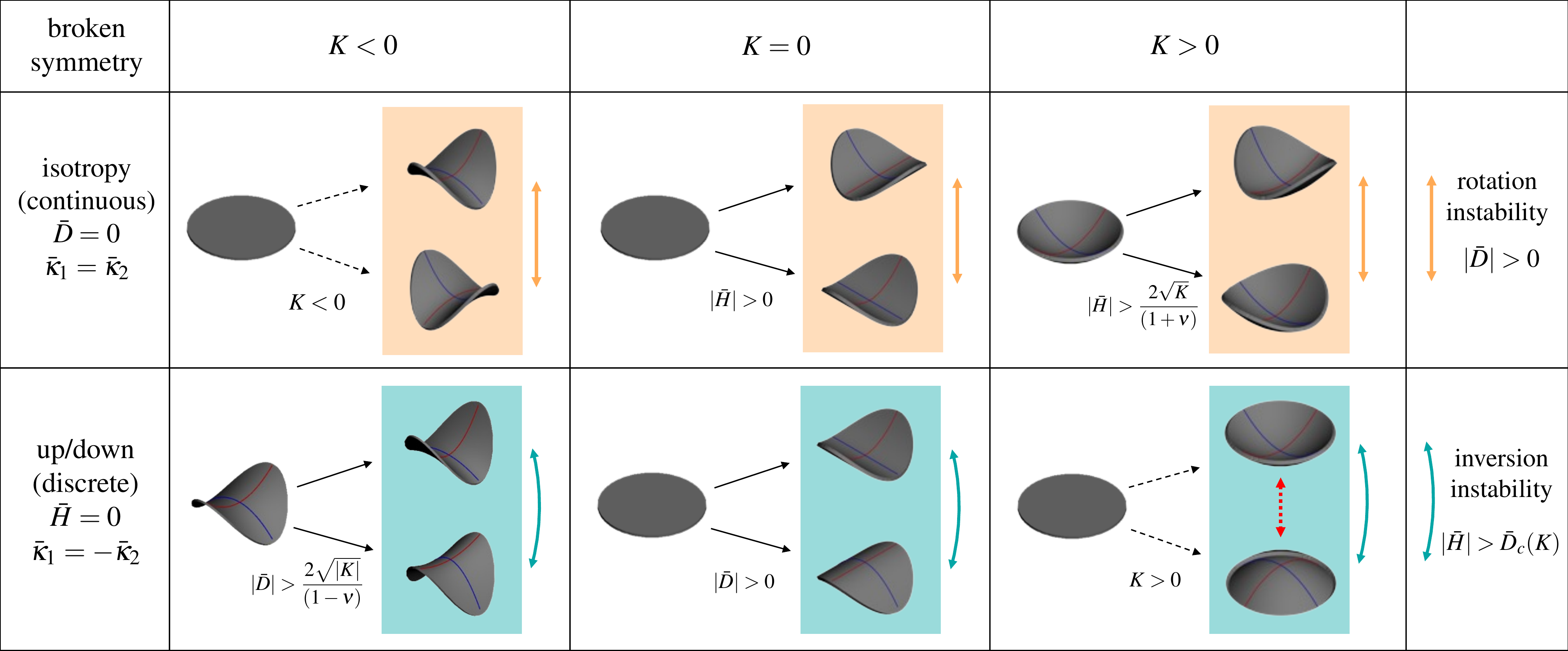}
    \caption{Summary of curvature induced instabilities in thin shallow shells with different Gauss curvature. Top row: shells with isotropic preferred curvature $\bD=0$ break isotropy, making  rotationally degenerate states (two shown) that are continuously connected. Isotropy is broken by all $K<0$ shells, but only past a threshold $\bH$  for $K>0$.  Yellow arrows mark the associated \textit{rotation} instability (between degenerate states) associated with traversing the $\bD=0$ line, with vanishing threshold, $\bD=0^{\pm}$.  Second row: shells with up-down symmetric preferred curvature, $\bH=0$, break this discrete symmetry leading to two equivalent states and bistability. Up-down symmetery is broken by all $K>0$ shells, but past a threshold $\bD$ for $K<0$.  The \textit{inversion} instability, marked by the curved cyan arrows, is associated with exiting the bistable region and jumping to an inverted state. The finite thresholds for inversion is eqn.\ \eqref{K0inversion_threshold} for $K=0$ and eqn.\ \eqref{bistablen} for $K<0$, while for $K>0$ inversion requires a non isometric pathway, leading to threshold that diverges in the thin limit.}
    \label{fig:instabplot}
\end{figure}

The resultant model has three cases delineated by Gauss curvature, $K<0$, $K=0$ and $K>0$. In each case, we have given a full phase diagram showing the achieved state of the shell as a function of the prefered curvature. The phase diagrams reveal three types of curvature driven instabilities, in which the principal achieved curvatures of the shell \textit{rotate}, \textit{symmetry-break} and \textit{invert},  as summarised in Fig.\ \ref{fig:instabplot}. The model predicts modest and thickness independent threshold-curvatures $\propto\sqrt{K}$ for all the instabilities, except inversion in the $K>0$ case, which cannot proceed isometrically, and hence has a thickness dependent threshold that diverges in the thin limit, and is beyond the reach of the simple model.

\textit{Symmetry-breaking} transitions arise in two flavours. In one case, we have isotropic preferred curvature $\bar{\kappa_1}=\bar{\kappa_2}$ ($\bD=0$) that is larger than permitted by the Gauss constraint, $\kappa_1\kappa_2=K$. It eventually  becomes favourable for one achieved curvature to grow at the expense of the other, to better fit the preferred curvature while maintaining $K$. For $K=0$ this occurs for any finite preferred curvature, and the system rolls up in an arbitrary direction, while for $K>0$ this involves a spherical cap folding along an arbitrary direction, and occurs super-critically past a threshold preferred curvature. For $K<0$ isotropy is always broken by the shell due to the Gauss requirement to form a saddle, even when the preferred curvature is actually zero, so this type of instability is not observed by changing preferred curvatures at fixed $K<0$. However, the symmetry broken state is present, and the instability  can be observed by an active change in the the metric from flat to $K<0$ \cite{Efrati2009}, in which case the saddle must ``decide" an orientation as it forms. Since this symmetry breaking is associated with breaking of a countinuous (rotationally invariant) symmetry, we call a \textit{continuous symmetry breaking} instability. 

The second \textit{symmetry-breaking} case arises with equal and opposite preferred curvatures, $\bar{\kappa_1}=-\bar{\kappa_2}$ ($\bH=0$) encoding a preferred symmetric saddle. In the $K=0$ case, the sheet will roll up in one of these two ways, while, for $K<0$ it will form a symmetric saddle for low preferred curvatures, and either of two asymmetric saddles beyond a threshold.  The energetic motivation is the same as the continuous case, but this time a discrete up-down symmetry is broken, yielding bistability. We thus call it a \textit{discrete symmetry breaking} instability. In the $K>0$ case, up-down symmetry is always broken by the Gauss requirement to have both curvatures with the same sign, but the instability can again be observed by an active change in the the metric from flat to $K>0$ \cite{Efrati2009}, in which case the cap must ``decide" whether to pop up or down as it forms. Curiously, breaking isotropy for $K>0$ and breaking up-down symmetry for $K<0$ are described by almost identical threshold and amplitude equations, except with $D$ and $H$ as the order parameter, and $\bH$ and $\bD$ as the control parameter.  

The buckling threshold of  \textit{symmetry breaking} instabilities in both negative and positive Gauss curved systems depends on the magnitude of the preferred curvature load relative to that of the Gauss curvature. Throughout, we have  discussed instabilities in terms of changing extrinsic preferred curvature, $\bar{\kappa}_1$ and $\bar{\kappa}_2$, at constant $K$, indicating active shape changes that vary through the thickness. However, one may achieve the same ends by varying $K$ at fixed $\bar{\kappa}_1$ and $\bar{\kappa}_2$ indicating active shape changes to the metric. In such a case, the shell will move along a diagonal line radiating out from the origin on any of the phase diagrams, which suffices to trigger either symmetry breaking transition by moving along the $\bD=0$ and $\bH=0$ axes respectively.

The \textit{rotation} instabilities arise as a consequence of states that break isotropy. This angular symmetry breaking leads to Mexican-hat style energies in which the curvature frame may rotate without penalty. Applying a small bias to such a potential, by traversing the $\bD=0$ line,  will break the degeneracy and produce a large rotation without an energy change. All three Gauss cases have such states, and hence show rotation instabilities. Finally, the \textit{inversion} instabilities arise from changing the sign of $\bH$ to apply an up-down bias to the bistability  generated by the \textit{discrete symmetry-breaking} transitions. In this case, a finite amount of $\bH$ is required to eliminate a minima, leading to a sub-critical inversion of the shell at the limit of bistability. For $K<0$ cases, this occurs at a modest thickness-independent threshold, while for $K>0$ it does not, so the instability is not present in the model, though it does occur in real shells via boundary layer stretching \cite{Pezzulla2018,Taffetani2018}.

Rotation and inversion instabilities are not naturally triggered by manipulating $K$ rather than $\bar{D}$ and $\bar{H}$, as they occur by crossing the $\bar{D}=0$ and $\bar{H}=0$ axes on the phase diagram, which no line radiating from the origin will do. However, for $K<0$, if a shell is in the local minima of the bistable region, it can exit the bistable region and invert via changes in $K$, as illustrated by the grey line in the phase diagram of Fig.\ \ref{fig:KNphase}.

The omission of stretch effects and boundary-layer mechanics underpins the simplicity of our model, but is also its key limitation. Although the boundary layer may be safely neglected in suitably thin shells, where the extent of the boundary layer, $\propto\sqrt{t}$, is negligible compared to the shell's extent, the boundary layer is of considerable importance in many physical shells. Firstly, the boundary layer permits the $K>0$ inversion instability. Moreover, even in very thin shells, the boundary layer can allow the shape of the shell's perimeter to couple its achieved curvature, particularly in the situations of a continuous broken symmetry; for example a Gauss flat bilayer square always rolls up along a diagonal \cite{seffen1999deployment, Pezzulla2017, Jiang2018}, rather than being truly degenerate. In modest thickness shells, the boundary layer can also nucleate or delay even those instabilities which could proceed isometrically \cite{Jiang2018,Pezzulla2018}. However, the simple isometric model nevertheless gives considerable insight into why these instabilities occur, and when the boundary layer is a complication rather than an essential ingredient. It also clarifies when we should expect thickness independent thresholds, and is asymptotically correct at small thickness. 

In contrast, the shallowness assumption in our model does not appear to be a major limitation. Indeed, we have shown that the threshold for the symmetry-breaking folding of a spherical cap subject to excessive preferred curvature is also correct for a deep spherical shell. Here the Rayleigh isometry of the sphere potentially allows a much deeper and richer exploration of the response of such a shell to excessive curvature, which we reserve for future work.

\vskip6pt

\enlargethispage{20pt}

\section*{Acknowledgement}
JSB is supported by a UKRI \emph{Future Leaders Fellowship}, grant number MR/S017186/1. AG thanks the EPSRC for funding, project 2108804.

\bibliographystyle{RS}
\bibliography{BendSnaps}

\begin{thebibliography}{99}

\bibitem{Euleroo}
Euler L. 1744 {\em Opera Omnia} vol.~I.

\bibitem{yoo2011stability}
Yoo CH, Lee S. 2011 {\em Stability of structures: principles and applications}.
Elsevier.

\bibitem{zoleypressure}
Zoley R. 1915 {\em Über ein Knickproblem an der Kugelschale}.
PhD thesis Zurich.

\bibitem{Hutchinson2016}
Hutchinson JW. 2016  {Buckling of spherical shells revisited}. {\em Proceedings
  of the Royal Society A: Mathematical, Physical and Engineering Sciences}
  \textbf{472}.

\bibitem{kochmann2017exploiting}
Kochmann DM, Bertoldi K. 2017  Exploiting microstructural instabilities in
  solids and structures: from metamaterials to structural transitions. {\em
  Applied mechanics reviews} \textbf{69}.

\bibitem{hirokawa1984volume}
Hirokawa Y, Tanaka T. 1984  Volume phase transition in a non-ionic gel. In {\em
  AIP Conference Proceedings} vol. 107 pp. 203--208. American Institute of
  Physics.

\bibitem{klein2007shaping}
Klein Y, Efrati E, Sharon E. 2007  Shaping of elastic sheets by prescription of
  non-Euclidean metrics. {\em Science} \textbf{315}, 1116--1120.

\bibitem{kim2012designing}
Kim J, Hanna JA, Byun M, Santangelo CD, Hayward RC. 2012  Designing responsive
  buckled surfaces by halftone gel lithography. {\em Science} \textbf{335},
  1201--1205.

\bibitem{na2016grayscale}
Na JH, Bende NP, Bae J, Santangelo CD, Hayward RC. 2016  Grayscale gel
  lithography for programmed buckling of non-Euclidean hydrogel plates. {\em
  Soft Matter} \textbf{12}, 4985--4990.

\bibitem{gladman2016biomimetic}
Gladman AS, Matsumoto EA, Nuzzo RG, Mahadevan L, Lewis JA. 2016  Biomimetic 4D
  printing. {\em Nature materials} \textbf{15}, 413--418.

\bibitem{thompson1942growth}
Thompson DW, Thompson DW. 1942 {\em On growth and form} vol.~2.
Cambridge university press Cambridge.

\bibitem{savin2011growth}
Savin T, Kurpios NA, Shyer AE, Florescu P, Liang H, Mahadevan L, Tabin CJ. 2011
   On the growth and form of the gut. {\em Nature} \textbf{476}, 57--62.

\bibitem{shyer2013villification}
Shyer AE, Tallinen T, Nerurkar NL, Wei Z, Gil ES, Kaplan DL, Tabin CJ,
  Mahadevan L. 2013  Villification: how the gut gets its villi. {\em Science}
  \textbf{342}, 212--218.

\bibitem{goriely2017mathematics}
Goriely A. 2017 {\em The mathematics and mechanics of biological growth}
  vol.~45.
Springer.

\bibitem{de2012engineering}
de~Haan LT, S{\'a}nchez-Somolinos C, Bastiaansen CM, Schenning AP, Broer DJ.
  2012  Engineering of complex order and the macroscopic deformation of liquid
  crystal polymer networks. {\em Angewandte Chemie International Edition}
  \textbf{51}, 12469--12472.

\bibitem{ware2015voxelated}
Ware TH, McConney ME, Wie JJ, Tondiglia VP, White TJ. 2015  Voxelated liquid
  crystal elastomers. {\em Science} \textbf{347}, 982--984.

\bibitem{aharoni2018universal}
Aharoni H, Xia Y, Zhang X, Kamien RD, Yang S. 2018  Universal inverse design of
  surfaces with thin nematic elastomer sheets. {\em Proceedings of the National
  Academy of Sciences} \textbf{115}, 7206--7211.

\bibitem{barnes2019direct}
Barnes M, Verduzco R. 2019  Direct shape programming of liquid crystal
  elastomers. {\em Soft matter} \textbf{15}, 870--879.

\bibitem{siefert2019bio}
Si{\'e}fert E, Reyssat E, Bico J, Roman B. 2019  Bio-inspired pneumatic
  shape-morphing elastomers. {\em Nature materials} \textbf{18}, 24--28.

\bibitem{warner2020inflationary}
Warner M, Siéfert E. 2020  Inflationary routes to Gaussian curved topography.
  {\em Proc. R. Soc. A} \textbf{476}, 20200047.

\bibitem{mosadegh2014pneumatic}
Mosadegh B, Polygerinos P, Keplinger C, Wennstedt S, Shepherd RF, Gupta U, Shim
  J, Bertoldi K, Walsh CJ, Whitesides GM. 2014  Pneumatic networks for soft
  robotics that actuate rapidly. {\em Advanced functional materials}
  \textbf{24}, 2163--2170.

\bibitem{bowden1998spontaneous}
Bowden N, Brittain S, Evans AG, Hutchinson JW, Whitesides GM. 1998  Spontaneous
  formation of ordered structures in thin films of metals supported on an
  elastomeric polymer. {\em nature} \textbf{393}, 146--149.

\bibitem{jiang2007finite}
Jiang H, Khang DY, Song J, Sun Y, Huang Y, Rogers JA. 2007  Finite deformation
  mechanics in buckled thin films on compliant supports. {\em Proceedings of
  the National Academy of Sciences} \textbf{104}, 15607--15612.

\bibitem{sultan2008buckling}
Sultan E, Boudaoud A. 2008  The buckling of a swollen thin gel layer bound to a
  compliant substrate. {\em Journal of applied mechanics} \textbf{75}.

\bibitem{huang2005nonlinear}
Huang Z, Hong W, Suo Z. 2005  Nonlinear analyses of wrinkles in a film bonded
  to a compliant substrate. {\em Journal of the Mechanics and Physics of
  Solids} \textbf{53}, 2101--2118.

\bibitem{audoly2007buckling}
Audoly B, Boudaoud A. 2007  Buckling of a thin film bound to a compliant
  substrate (part I). {\em Formulation, linear stability of cylindrical
  patterns, secondary bifurcations. Submitted to Journal of the Mechanics and
  Physics of Solids}.

\bibitem{brain}
Tallinen T, Chung JY, Rosseau F, Girard N, Lefevre J, Mahadevan L. 2016  On the
  growth and form of cortical convolutions. {\em Nature Phys.} \textbf{12},
  588--593.

\bibitem{brainpnas}
Tallinen T, Chung JY, Biggins JS, Mahadevan L. 2014  Gyrification from
  constrained cortical expansion. {\em PNAS} \textbf{111}, 12667--12672.

\bibitem{villi}
Shyer AE, Tallinen T, Nerurkar NL, Wei Z, Gil ES, Kaplan. DL, Tabin CJ,
  Mahadevan L. 2013  Villification: How the Gut Gets Its Villi. {\em Science}
  \textbf{342}, 212--218.

\bibitem{loop}
Savin T, Kurpios NA, Shyer AE, Florescu P, Liang H, Mahadevan L, Tabin CJ. 2011
   On the growth and form of the gut. {\em Nature} \textbf{476}, 57--63.

\bibitem{forterre2005venus}
Forterre Y, Skotheim JM, Dumais J, Mahadevan L. 2005  How the Venus flytrap
  snaps. {\em Nature} \textbf{433}, 421--425.

\bibitem{Katifori2010}
Katifori E, Alben S, Cerda E, Nelson DR, Dumais J. 2010  {Foldable structures
  and the natural design of pollen grains}. {\em Proceedings of the National
  Academy of Sciences of the United States of America} \textbf{107},
  7635--7639.

\bibitem{Couturier2013}
Couturier E, Dumais J, Cerda E, Katifori E. 2013  {Folding of an opened
  spherical shell}. {\em Soft Matter} \textbf{9}, 8359--8367.

\bibitem{Bozic2020}
Bo{\v{z}}i{\v{c}} A, {\v{S}}iber A. 2020  {Mechanical design of apertures and
  the infolding of pollen grain}. {\em Proceedings of the National Academy of
  Sciences of the United States of America} \textbf{117}, 26600--26607.

\bibitem{seffen1999deployment}
Seffen K, Pellegrino S. 1999  Deployment dynamics of tape springs. {\em
  Proceedings of the Royal Society of London. Series A: Mathematical, Physical
  and Engineering Sciences} \textbf{455}, 1003--1048.

\bibitem{kuhnel2015differential}
K{\"u}hnel W. 2015 {\em Differential geometry} vol.~77.
American Mathematical Soc.

\bibitem{van2017growth}
van Rees WM, Vouga E, Mahadevan L. 2017  Growth patterns for shape-shifting
  elastic bilayers. {\em Proceedings of the National Academy of Sciences}
  \textbf{114}, 11597--11602.

\bibitem{warner2020topographic}
Warner M. 2020  Topographic mechanics and applications of liquid crystalline
  solids. {\em Annual Review of Condensed Matter Physics} \textbf{11},
  125--145.

\bibitem{timoshenko1925analysis}
Timoshenko S. 1925  Analysis of bi-metal thermostats. {\em Josa} \textbf{11},
  233--255.

\bibitem{Efrati2009}
Efrati E, Sharon E, Kupferman R. 2009  {Elastic theory of unconstrained
  non-Euclidean plates}. {\em Journal of the Mechanics and Physics of Solids}
  \textbf{57}, 762--775.

\bibitem{Pezzulla2018}
Pezzulla M, Stoop N, Steranka MP, Bade AJ, Holmes DP. 2018  {Curvature-Induced
  Instabilities of Shells}. {\em Physical Review Letters} \textbf{120}, 48002.

\bibitem{pandey2014dynamics}
Pandey A, Moulton DE, Vella D, Holmes DP. 2014  Dynamics of snapping beams and
  jumping poppers. {\em EPL (Europhysics Letters)} \textbf{105}, 24001.

\bibitem{mansfield1989bending}
Mansfield EH. 1989 {\em The bending and stretching of plates}.

\bibitem{kebadze2004bistable}
Kebadze E, Guest S, Pellegrino S. 2004  Bistable prestressed shell structures.
  {\em International Journal of Solids and Structures} \textbf{41}, 2801--2820.

\bibitem{Pezzulla2016}
Pezzulla M, Smith GP, Nardinocchi P, Holmes DP. 2016  {Geometry and mechanics
  of thin growing bilayers}. {\em Soft Matter} \textbf{12}, 4435--4442.

\bibitem{Pezzulla2017}
Pezzulla M, Stoop N, Jiang X, Holmes DP. 2017  {Curvature-driven morphing of
  non-Euclidean shells Subject Areas :}. {\em Proceedings of the Royal Society
  A: Mathematical, Physical and Engineering Sciences}.

\bibitem{Jiang2018}
Jiang X, Pezzulla M, Shao H, Ghosh TK, Holmes DP. 2018  {Snapping of bistable,
  prestressed cylindrical shells}. {\em Epl} \textbf{122}.

\bibitem{sawa2011shape}
Sawa Y, Ye F, Urayama K, Takigawa T, Gimenez-Pinto V, Selinger RL, Selinger JV.
  2011  Shape selection of twist-nematic-elastomer ribbons. {\em Proceedings of
  the National Academy of Sciences} \textbf{108}, 6364--6368.

\bibitem{efrati2014orientation}
Efrati E, Irvine WT. 2014  Orientation-dependent handedness and chiral design.
  {\em Physical Review X} \textbf{4}, 011003.

\bibitem{gao2021molecularly}
Gao J, Clement A, Tabrizi M, Shankar MR. 2021  Molecularly Directed,
  Geometrically Latched, Impulsive Actuation Powers Sub-Gram Scale Motility.
  {\em Advanced Materials Technologies} p. 2100979.

\bibitem{Taffetani2018}
Taffetani M, Jiang X, Holmes DP, Vella D. 2018  {Static bistability of
  spherical caps}. {\em arXiv}.

\bibitem{strutt1877theory}
Strutt JW, Rayleigh JWSB. 1894 {\em The theory of sound} vol.~1.
Macmillan.

\end{thebibliography}

\end{document}